\documentclass{aastex62}
\usepackage{enumerate}

\shorttitle{IFS observations of the ionized gas in M81}
\shortauthors{Li et al.}

\usepackage{graphicx}	
\usepackage{amsmath}	
\usepackage{amssymb}	

\usepackage[normal]{threeparttable}

\begin{document}

\title{CAHA/PPAK Integral-field Spectroscopic Observations of M81. I. Circumnuclear ionized gas}
\author{Zongnan Li}
\email{lzn@smail.nju.edu.cn}
\author{Zhiyuan Li}
\email{lizy@nju.edu.cn}
\affiliation{School of Astronomy and Space Science, Nanjing University, Nanjing 210023, China}
\affiliation{Key Laboratory of Modern Astronomy and Astrophysics, Nanjing University, Nanjing 210023, China}
\author{Rub\'{e}n Garc\'{\i}a-Benito}
\affiliation{Instituto de Astrof\'{\i}sica de Andaluc\'{\i}a (CSIC), P.O. Box 3004, 18080 Granada, Spain}
\author[0000-0002-9767-9237]{Shuai Feng}
\affiliation{College of Physics, Hebei Normal University, 20 South Erhuan Road, Shijiazhuang, 050024, China}
\affiliation{Hebei Key Laboratory of Photophysics Research and Application, 050024 Shijiazhuang, China}

\begin{abstract}

Galactic circumnuclear environments of nearby galaxies provide unique opportunities for our understanding of the co-evolution between super-massive black holes and their host galaxies. Here we present a detailed study of ionized gas in the central kiloparsec region of M81, which hosts the closest prototype low-luminosity active galactic nucleus, based on optical integral-field spectroscopic observations taken with the CAHA 3.5m telescope. 
It is found that much of the circumnuclear ionized gas is concentraed within a bright core of $\sim$200 pc in extent and a surrounding spiral-like structure known as the nuclear spiral. 
The total mass of the ionized gas is estimated to be $\sim2\times10^5\rm~M_\odot$, which corresponds to a few percent of the cold gas mass in this region, as traced by co-spatial dust extinction features. 
Plausible signature of a bi-conical outflow along the disk plane is suggested by a pair of blueshifted/redshifted low-velocity features, symmetrically located at $\sim$ 120 -- 250 pc from the nucleus. The spatially-resolved line ratios of [N\,{\sc ii}]/H$\alpha$ and [O\,{\sc iii}]/H$\beta$ demonstrate that much of the circumnuclear region can be classified as LINER (low-ionization nuclear emission-line region). 
However, substantial spatial variations in the line intensities and line ratios strongly suggest that different ionization/excitation mechanisms, rather than just a central dominant source of photoionization, are simultaneously at work to produce the observed line signatures. 
\end{abstract}


\section{Introduction}

Galactic circumnuclear environments, in which the multi-phase interstellar medium (ISM) and various stellar populations are coupled under the strong influence of a super-massive black hole (SMBH), if present, are crucial for our understanding of the co-evolution of SMBHs and their host galaxies. 
The circumnuclear ISM, itself supplied by externally acquired material and/or local stellar ejecta, is the necessary fuel for the SMBH to become an active galactic nucleus (AGN).
Meanwhile, the circumnuclear ISM also provides sensitive probes to nuclear activities, chiefly through their spectroscopic and kinematic behavior. 
\\
 
The recent advent of integral-field spectroscopy (IFS) surveys have enabled systematic studies of nearby galaxies based on large sets of optical spectra, providing new insights in particular to the circumnuclear environments.  
Low ionization nuclear emission-line regions (LINERs; \citealp{1980A&A....87..152H}), which are characterized by prominent low-excitation emission lines of weakly ionized atoms, have been identified in a large number of inactive galaxies by the pioneer Spectrographic Areal Unit for Research on Optical Nebulae (SAURON) survey \citep{2010MNRAS.402.2187S}, and later by the Calar Alto Legacy Integral Field spectroscopy Area \citep[CALIFA;][]{2012A&A...538A...8S}, Sydney-AAO Multi-object Integral-field spectrograph \citep[SAMI;][]{2015MNRAS.447.2857B} and Mapping Nearby Galaxies at Apache Point Observatory (MaNGA) surveys \citep{2016MNRAS.461.3111B}. A variety of physical mechanisms have long been proposed for the ionization and excitation of LINERs, which include photoionization by a low-luminosity AGN \citep[LLAGN;][]{1983ApJ...264..105F, 1983ApJ...269L..37H}, massive young stars (e.g., OB stars and Wolf-Rayet stars; \citealp{1992ApJ...397L..79F, 2000PASP..112..753B}), hot evolved stars (e.g., planetary nebulae and post-asymptotic giant branch [pAGB] stars;  \citealp{1994A&A...292...13B, 2008MNRAS.391L..29S, 2012ApJ...747...61Y, 2013A&A...558A..43S}), as well as shocks \citep{1980A&A....87..152H, 1997ApJ...490..202D} and cosmic rays \citep{1984ApJ...286...42F}. 
The leading role of AGN photoionization in LINERs is revisited by the IFS observations and often called into question due to insufficient ionizing photons from a LLAGN \citep{2010ApJ...711..796E, 2016MNRAS.461.3111B}. 
More recently, higher-resolution Gemini Multi-Object Spectrograph (GMOS) IFS observations and $Hubble~Space~Telescope$ (HST) slit observations \citep{2015MNRAS.451.3728R, Molina_2018} of a small samples of LINERs also suggest that radiation from the central LLAGN is short for the ionization, and that other mechanisms, such as pAGB stars and shocks induced by LLAGN-driven jets/outflows should play a significant role. 
In addition, gas motions deviating from a pure rotating disk are discovered in LINERs, both by HST observations \citep{2008AJ....136.1677W} and by MaNGA on more extended regions \citep{2019MNRAS.486..344R}, suggesting the presence of outflows. 
Clearly, a spatially-resolved view of the spectroscopic and kinematic properties of the circumnuclear ionized gas is key to understanding the universality and relative contributions of various ionization/excitation mechanisms. \\

At a distance of 3.6 Mpc (1$\arcsec$ corresponds to 17.5 pc; \citealp{2001ApJ...553...47F}), M81 hosts the nearest prototype LLAGN, known as M81* \citep{1999ApJ...516..672H}, which has a bolometric luminosity of 2 $\times$ 10$^{41}$ erg s$^{-1}$, corresponding to an Eddington ratio of 3 $\times$ 10$^{-5}$ \citep{2014MNRAS.438.2804N} for an estimated SMBH mass of 7$\times 10^7$ M$_\odot$ \citep{2003AJ....125.1226D}.  Thanks to its proximity, the central region of M81 has been extensively investigated in multi-wavelengths and classified as both LINER \citep{1980A&A....87..152H} and type-1 Seyfert \citep{1981ApJ...245..845P}, the property of which was reviewed in \cite{1996ApJ...462..183H}. 
AGN activities are evidenced by a variable, power-law X-ray continuum \citep{1996PASJ...48..237I}, a featureless, compact UV continuum \citep{1996ApJ...462..183H, 1997ApJ...481L..71D}, optical double-peaked broad emission lines \citep{1996AJ....111.1901B}, and a compact radio core plus a one-sided, precessing jet \citep{2001MNRAS.321..767I, 2011A&A...533A.111M}. M81 thus provides us with one of the best laboratories for studying the interplay between an LLAGN and its close environment.
\\

M81* is found to interact vividly with the its environment on various scales.
GMOS IFS observations have found signatures of both an inflow of ionized gas at a rate of $3\times 10^{-3}$ M$_\odot$ yr$^{-1}$ and an outflow vertical to a rotating ionized gas disk in the central 120 pc$\times 250$ pc region \citep{2011MNRAS.413..149S}. This outflow was reinterpreted by \cite{2015A&A...576A..58R} as in line with the radio jet. 
More recently, a high-velocity outflow of hot gas is detected in the central $\lesssim$ 40 pc region with $Chandra$ X-ray spectroscopy \citep{2021NatAs...5..928S}, which is most likely driven by the hot accretion flow onto the SMBH. 
Evidence for this hot wind interacting with the circumnuclear ISM was also suggested from detection of diffuse hot gas.
On larger scales, M81 resides in a well known galaxy group, where the main members, M81, M82 and NGC 3077 are actively interacting with each other. High-resolution, wide-field HI observations have found that these galaxies share a common atomic gas envelop \citep{1994Natur.372..530Y, 2018ApJ...865...26D}, and are connected with massive HI filaments. Arguably undergoing a merger with its satellite M82, M81 is fueled by tidally disrupted materials from both M82 and NGC 3077 \citep{2020ApJ...905...60S}. Such interactions could have significantly affected the nuclear activity in M81 in the recent past (or may do so in the near future).
\\

In the optical band, an H$\alpha$-bright gaseous structure across the central few hundred parsecs is the most conspicuous feature in the circumnuclear region of M81 \citep{1995AJ....110.1115D}. This structure has an  organized morphology characterized by spiral-like filaments (hence dubbed a {\it nuclear spiral}) extending into the close vicinity of the SMBH, raising the interesting question of whether it could be responsible for fueling the AGN activity. 
The ionization mechanism of this structure remains uncertain, although massive stars can be safely ruled out due to their current absence in this region \citep{1997ApJ...481L..71D}. The LLAGN M81* is a promising source of photoionization, especially for the innermost region \citep{1996ApJ...462..183H}. However, it remains to quantify the role of the LLAGN in outer regions and to examine possible contributions from other mechanisms mentioned in the above.  
In this regard, the case of M81 can be highly instructive when compared to the mysterious case of M31, which hosts the nearest known LINER in the clear absence of either an LLAGN or massive stars \citep{2009MNRAS.397..148L}.
\\

Although high-resolution spectroscopic observations, including those afforded by GMOS and HST \citep[e.g.,][]{2011MNRAS.413..149S, 2003AJ....125.1226D, 2019MNRAS.488.1199D}, have been carried out to resolve the properties of ionized gas in the inner $\sim$100 pc down to the central 1 pc of M81, sensitive IFS observations covering a wider area would still be critical to reveal the full connection between the SMBH and its circumnuclear environment. 
In particular, IFS observations hold promise for understanding the ionization mechanism(s) of the nuclear spiral in M81 through emission line diagnostics, which would help to resolve the enigma of LINERs in general. 
\\

We are thus motivated to carry out the first optical IFS observations with a full coverage of the central kpc region of M81, utilizing the same instrument and a similar observational setup as the CALIFA survey, which satisfy the above requirements. 
In this first paper of a series, we provide an overview of the observations and present the properties of the circumnuclear ionized gas as traced by their major emission lines. The properties of the stellar population, and a detailed investigation of the gas ionization mechanism will be presented in subsequent papers. In Section \ref{sec:2} we describe the observations and data reduction procedure. The morphology and kinematics of the ionized gas derived from the emission lines, as well as the gas properties inferred from the line ratios are presented in Section \ref{sec:3}. 
Some interesting implications of the observational results are addressed in Section \ref{sec:4}, followed by a summary in Section~\ref{sec:sum}.
\\

\section{Observation and data reduction} \label{sec:2}

The spectroscopic observations of M81 were taken with the PPAK integral field unit (IFU) of the Potsdam Multi-Aperture Spectrograph (PMAS) instrument at the 3.5 m telescope of the Centro Astron$\rm\acute{o}$mico Hispano Alem$\rm\acute{a}$n (CAHA) at Calar Alto. The observations were carried out on 2017-11-19 using the medium-resolution V1200 grating ($3650-4840\,\rm\mathring{A}$; FWHM 2.3 \AA) and on 2018-4-19 using the low-resolution V500 grating ($3745-7500\,\rm\mathring{A}$; FWHM 6 \AA), respectively. 
These settings are equivalent to those of the CALIFA survey.
\\

\begin{figure}
\centering
\includegraphics[width= 3.2in]{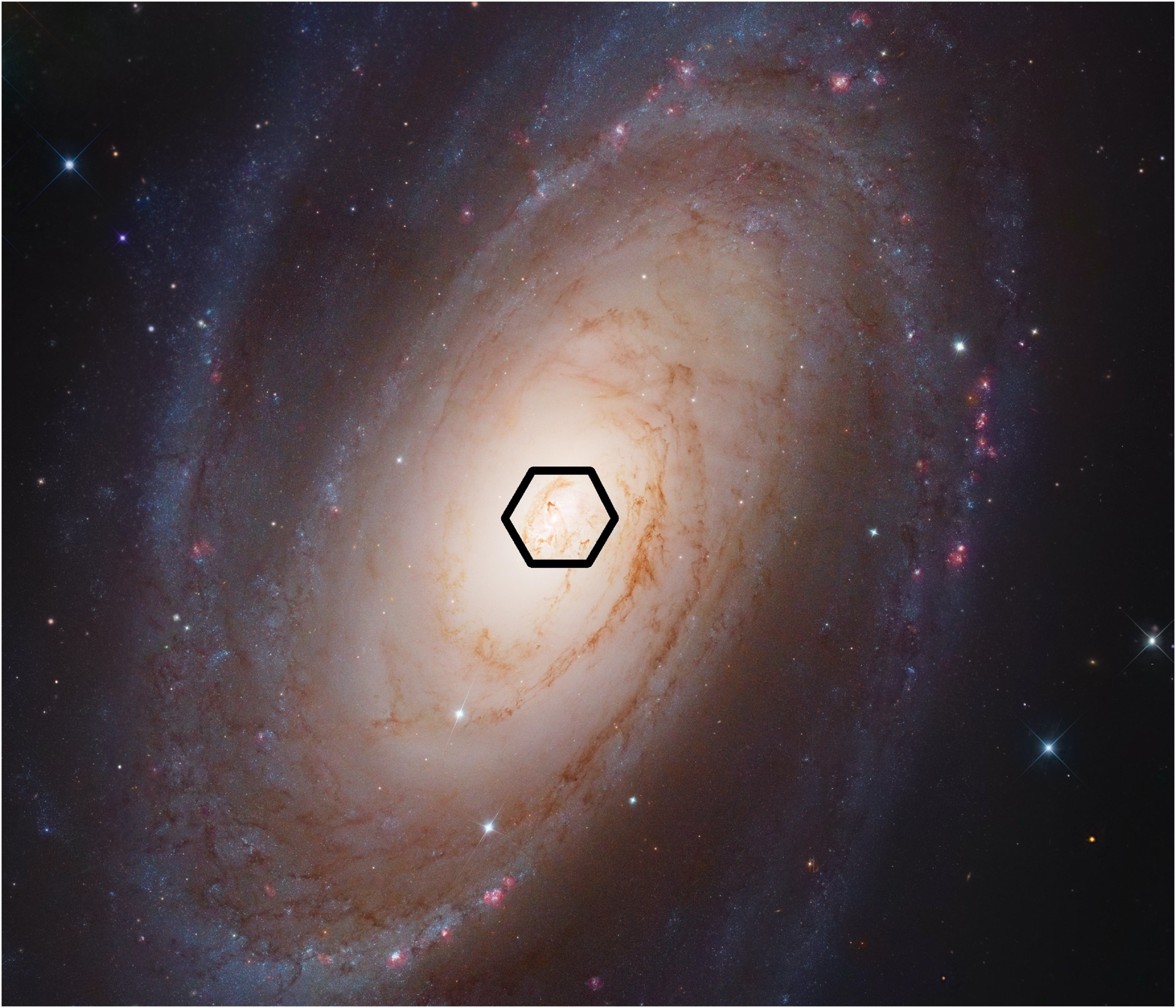}
\includegraphics[width= 3in]{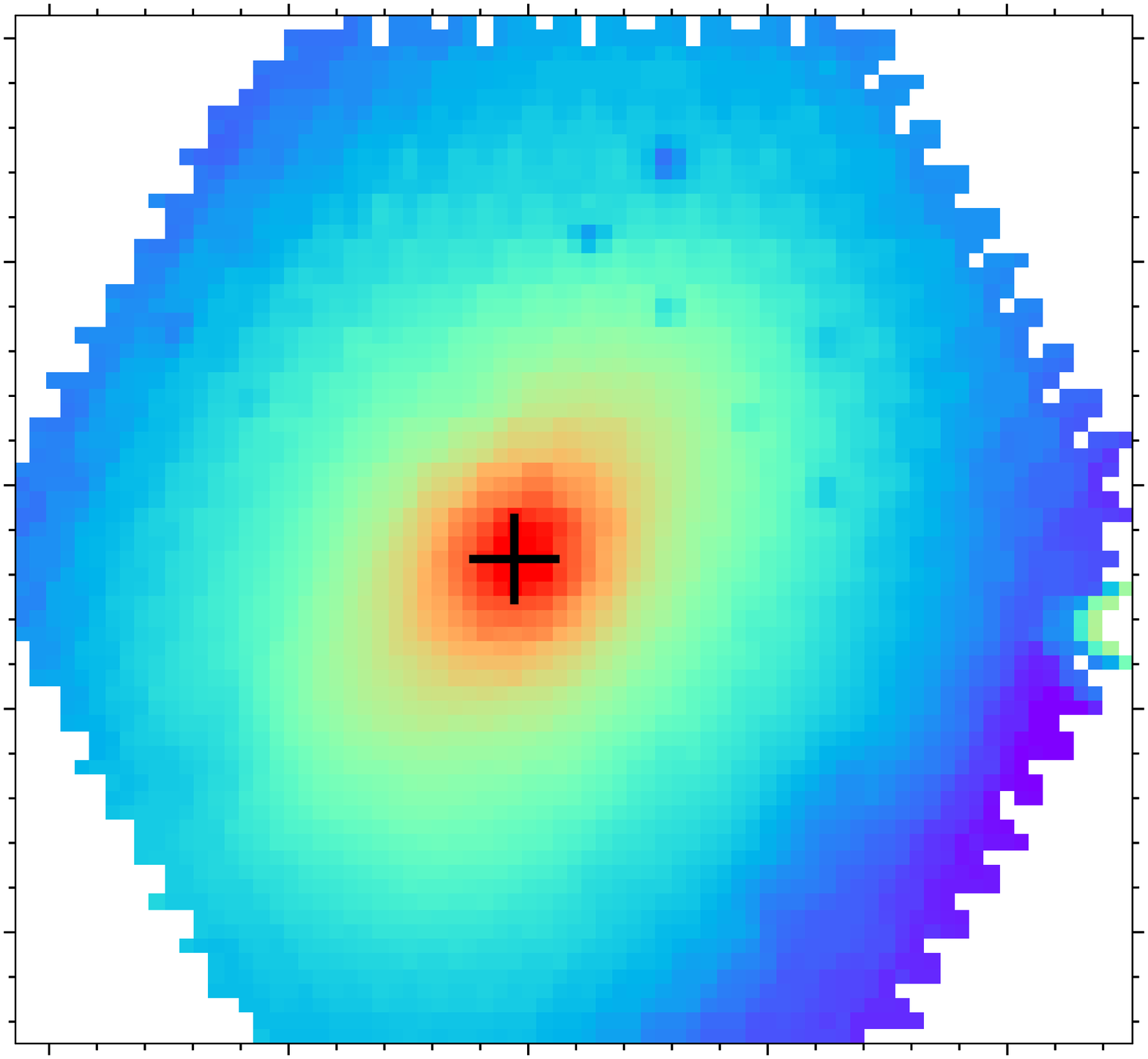}
\caption{$Left$: The field-of-view (74$\arcsec~\times$ 64$\arcsec$, black hexagon) of the PPAK observations overlaid on a composite HST and Subaru image of M81. 
$Right$: The continuum image integrated over the range of 3700--7300 $\rm\mathring{A}$ from the PPAK observations. The nucleus of M81, marked by the black cross, is slightly offset toward southeast from the image center. \label{fig:fov}}
\end{figure}

A combined spectral cube (called COMBO) of the two gratings were produced after convolving the V1200 resolution to match that of the V500. The COMBO cube solves the problem of vignetting that otherwise affects the blue end of the V500. 
The final wavelength coverage ranges from 3700 to 7300 $\rm\mathring{A}$, 
corrected for the heliocentric velocity and vignetting at the spectral edges.
The field-of-view (FoV) is approximately a hexagon of 74$\arcsec~\times$ 64$\arcsec$, sampled by $2.7\arcsec$ fibers.
A three-pointing dithering scheme was adopted to ensure a full FoV coverage of the spectral cube, which was resampled to have a pixel size of $1\arcsec$ ($\sim$ 17.5 pc at the assumed distance of M81). We carried out V1200 observations with an exposure time of 3600 s per pointing (split into 3 individual exposures), while for the V500 setup we obtained observations with a total of 2100 s per pointing (split into 7 individual exposures to avoid saturation). 
The FoV is illustrated in Figure \ref{fig:fov} against an optical image of M81 combining HST and Subaru Telescope observations\footnote{Copyright 2016 Robert Gendler, Roberto Colombari, Hubble Legacy Archive, Subaru Telescope}. 
\\

Data reduction procedure followed the standard automatic pipeline of CALIFA \citep{2012A&A...538A...8S}. The main steps are summarized below: 1) CCD frame alignment and cosmic-ray removal; 2) Spectral extraction; 3) Sky subtraction; 4) Flux calibration; 5) Spatial re-arrangement and image reconstruction; 6) Atmospheric refraction and extinction correction. 
Both atmospheric and Galactic extinction have been corrected, which is 0.14 mag and 0.25 mag in the V-band, respectively. 
For more details of the data reduction procedure readers are referred to \citet{2013A&A...549A..87H}, \citet{2015A&A...576A.135G} and \citet{2016A&A...594A..36S}. 
An absolute flux calibration was not performed, as the flux accuracy is empirically determined to be within $\sim8\%$ with respect to the Sloan Digital Sky Survey \citep{2012A&A...538A...8S}.
\\

To determine the properties (e.g., line-of-sight velocity, velocity dispersion, and metallicity) of the stellar populations, the observed spectra were fitted pixel-by-pixel with the {\sc starlight} software \citep{2005MNRAS.358..363C, 2011ascl.soft08006C}, which utilizes synthesized spectra of single stellar populations (SSPs), following \cite{2017A&A...608A..27G}. The results were then processed through PyCASSO (the Python CALIFA starlight Synthesis Organizer; \citealp{2013A&A...557A..86C, 2017MNRAS.471.3727D}) to produce a suite of spatially resolved stellar population properties. 
Expected significant emission lines, as well as bad pixels, were masked during the fitting. The stellar population synthesis model consists of 254 SSPs from the GRANADA library \citep[][for stellar populations younger than 60 Myr]{2005MNRAS.357..945G} and \cite{2015MNRAS.449.1177V} based on BaSTi isochrones (for older ages). The metallicity ranges from -2.3 to +0.4 (with 8 values, log $Z/Z_\odot = $ -2.28, -1.79, -1.26, -0.66, -0.35, -0.06, +0.25 and +0.40), and the age ranges from 0.001 to 14 Gyr. A Salpeter initial mass function \citep[IMF;][]{1955ApJ...121..161S} was assumed here. A detailed investigation of the stellar populations will be presented in a future work (R. Garcia-Benito et al. in preparation).
\\

\begin{figure}
\centering
\includegraphics[width=7in]{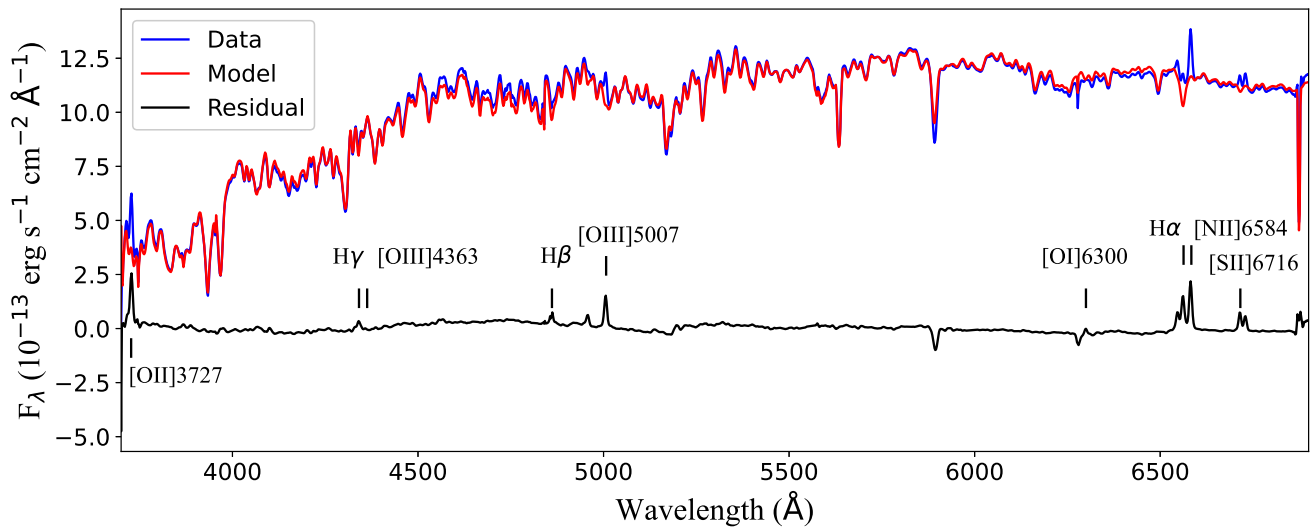}
\caption{The observed spectrum (blue curve) and the fitted stellar population model (red curve) stacked over all pixels. 
The emission line spectrum  (black curve) is obtained by subtracting the model from the observed spectrum. Prominent emission lines are labelled. \label{fig:eml}}
\end{figure}

Figure \ref{fig:eml} displays the observed spectrum accumulated over all pixels (blue curve), the cumulative modelled stellar spectrum (red curve) and the model-subtracted residual spectrum (black curve). The latter consists primarily of significant emission lines, 
including [O\,{\sc ii}]$\lambda$3727, H$\gamma$($\lambda$4341), [O\,{\sc iii}]$\lambda$4363, H$\beta$($\lambda$4861), [O\,{\sc iii}]$\lambda$4959,5007 doublet, [O\,{\sc i}]$\lambda$6300, H$\alpha$($\lambda$6563), [N\,{\sc ii}]$\lambda$6548,6584 doublet and [S\,{\sc ii}]$\lambda$6716,6731 doublet.
Absorption features near 5895 \AA\ and 6290 \AA\ are due to imperfect subtraction of the stellar continuum or telluric line.
On a pixel-by-pixel basis, we derived the emission line parameters, including line flux, line centroid (line-of-sight velocity) and width (velocity dispersion) by fitting each line with a single Gaussian profile. The velocity dispersion was corrected for the instrumental broadening. 
The noise level varies among different lines and different positions; for reference, the median surface brightness value with signal-to-noise ratio ($S/N$) $\sim$3 for the H$\alpha$ line is 2.4 $\times 10^{-17}\rm~erg~s^{-1}~cm^{-2}~arcsec^{-2}$. 
\\

We employ ancillary images from the HST archive to assist our analysis: 1) a Wide Field Planetary Camera 2 (WFPC2) H$\alpha$ map of the central $\sim$1 arcmin region of M81,
originally presented by \citet{1997ApJ...481L..71D}, and 2) a Wide Field Camera 3 (WFC3) F438W image covering the central $160\arcsec \times 160\arcsec$ of M81 (project ID 11421), chiefly to quantify the starlight distribution and to reveal dust extinction features co-spatial with the ionized gas.
\\

\section{Analysis and Results} \label{sec:3}
\subsection{Morphology} \label{sec:3.1}

Unless otherwise stated, in the following we will focus on the four familiar strong emission lines, H$\alpha$, H$\beta$, [N\,{\sc ii}]$\lambda$6584 and [O\,{\sc iii}]$\lambda$5007, which 
conventionally provide important diagnostics of the ionized gas.
A $S/N$ $> 5$ cut is applied to each of these lines, except for the relatively weak H$\beta$, for which a lower $S/N$ cut $> 3$ is adopted.
\\

\begin{figure}
\centering
\includegraphics[width= 7in]{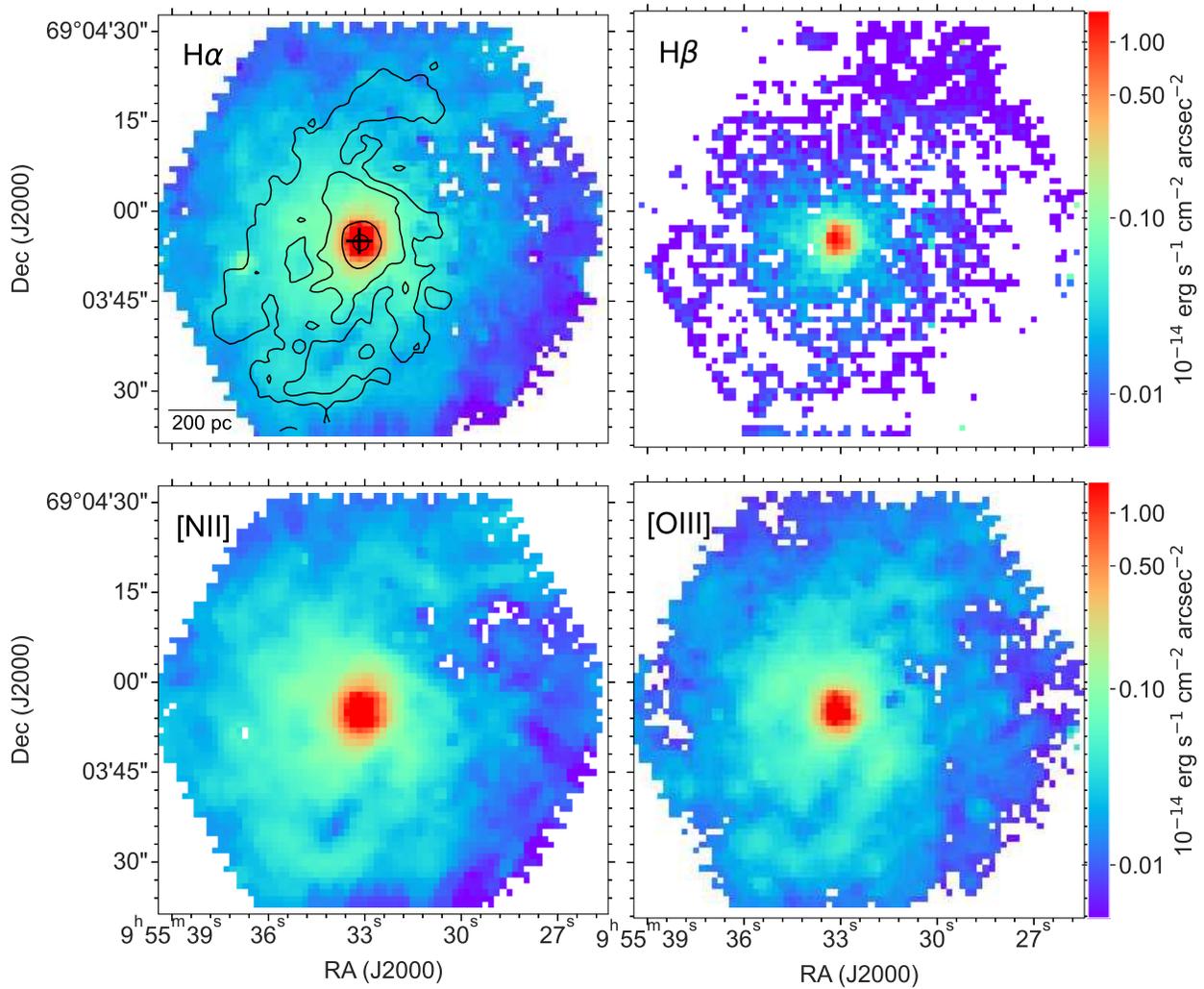}
\caption{Surface brightness distribution of H$\alpha$, H$\beta$, [N\,{\sc ii}]$\lambda$6584 and [O\,{\sc iii}]$\lambda$5007. 
A cut of $S/N > 3$ is applied to H$\beta$, while a cut of $S/N > 5$ is applied for the other three lines. The H$\alpha$ map is overlaid with H$\alpha$ intensity contours derived from the HST narrow-band image of \citet{1997ApJ...481L..71D}. 
Position of the nucleus is marked by a black cross.
\label{fig:flux}}
\end{figure}

\begin{figure}
\centering
\includegraphics[width=7in]{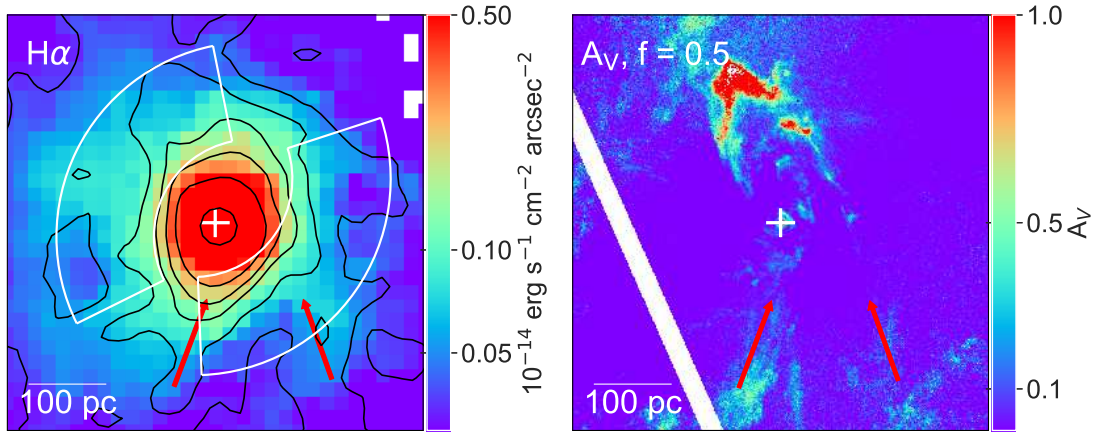}
\caption{$Left$: A zoom-in view of the H$\alpha$ surface brightness distribution in the central $30\arcsec \times 30\arcsec$, overlaid with HST H$\alpha$ contours (same as in Figure~\ref{fig:flux}). The two fan-shaped regions mark a possible outflow as revealed in the H$\alpha$ velocity field (Figure~\ref{fig:v}), and the red arrows mark the two arm-like features connecting with the core. $Right$: The same region in dust extinction of the $f = 0.5$ case, highlighting the dust lanes close to the nucleus, marked by a cross.
\label{fig:zoom-in}}
\end{figure}

The surface brightness distribution of the four lines are presented in Figure \ref{fig:flux}. 
All four lines show a highly similar morphology, although H$\beta$ appears patchy due to its relative dimness. 
The morphology is characterized by a bright core (within the central $\sim$200 pc; see a zoom-in view in Figure~\ref{fig:zoom-in}) surrounded by a lopsided, spiral-like structure with an extent of $\sim$ 1 kpc. This spiral structure, most prominent to the northeast and southwest of the core, was already reported by narrow-band imaging and dubbed the {\it nuclear spiral} \citep{1995AJ....110.1115D}. 
We show below that this prominent structure is also conspicuous in dust extinction. 
The H$\alpha$ map in Figure \ref{fig:flux} is contrasted with intensity contours derived from the HST H$\alpha$ image of \citet{1997ApJ...481L..71D}, which delineate the nuclear spiral at a higher resolution. 
Nevertheless, our H$\alpha$ map reveals the presence of more diffuse ionized gas across almost the entire FoV, which is not obvious in the HST narrow-band image.
We note that two additional strong emission lines, [O\,{\sc ii}] and [S\,{\sc ii}], share a similar morphology with H$\alpha$. 
For reference, the intensity maps of [O\,{\sc ii}] and [S\,{\sc ii}] are displayed in Figure \ref{fig:A1}. 
\\

\begin{figure}
\centering
\includegraphics[width=7in]{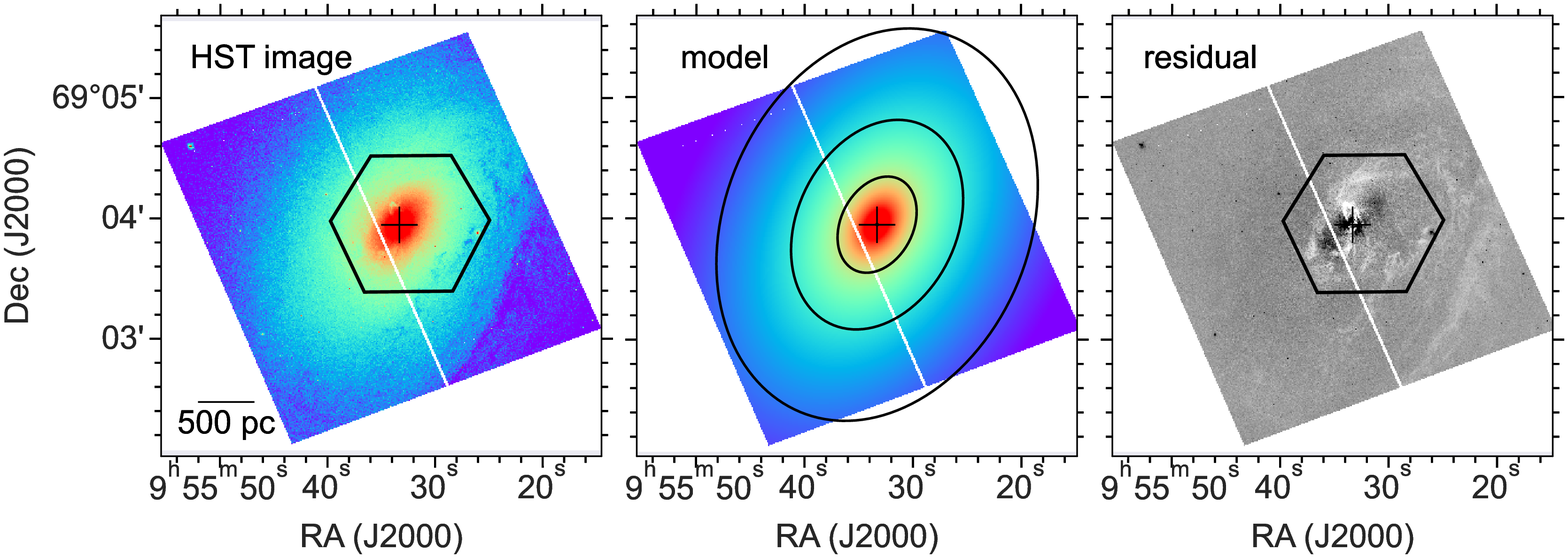}
\caption{$Left$: The HST WFC3/F438W image covering the inner region of M81. The hexagon delineates the PPAK FoV, and the black cross marks the M81 nucleus. $Middle$: The best-fit {\sc galfit} model composed of a bulge, a disk and a point-like nucleus. The contours indicate the bulge-to-total flux ratio, at levels of 80\%, 60\% and 40\% from inside out. $Right$: The residual image after subtracting the model image, in which dust extinction features stand out with negative values (represented by white color).
\label{fig:galfit}}
\end{figure}

The high-resolution dust extinction map shown in the right panel of Figure \ref{fig:galfit} is produced based on the HST WFC3/F438W image (left panel of Figure \ref{fig:galfit}), which covers a region a few times larger than the PPAK FoV.  
This wide-band, blue image is chosen for its sensitivity to uncovering dust extinction features against the bulge starlight.
We utilize the imaging fitting algorithm {\sc galfit} \citep[version 3.0.5,][]{Peng_2002, Peng_2010} to model the intrinsic starlight distribution,  
adopting three spatially-distinct physical components: a bulge, a disk, and a point source standing for the LLAGN M81*. The point-spread function (PSF) file suitable for the WFC/F438W image is downloaded from the HST website\footnote{https://www.stsci.edu/hst/instrumentation/wfc3/data-analysis/psf} and fed to {\sc galfit}, along with a mask image that accounts for bad pixels and bright, off-nuclear sources. \\

We fix the galactic center at the brightest pixel in the image, and allow the bulge S$\rm \acute{e}$rsic index $n$ (=4 for classical bulge, 1 for exponential disk; \citealp{1963BAAA....6...41S}) to vary during the fit. 
The resultant parameters of the three components are listed in Table \ref{tab1}, and the modeled image combining the three components is shown in the middle panel of Figure \ref{fig:galfit}. 
The S$\rm \acute{e}$rsic index $n$ derived from our fit is 3.23 $\pm$ 0.01,  indicating a classic bulge. 
The corresponding effective radius is 999 $\pm$ 5 pc, which is roughly covered by the PPAK FoV and well within the WFC3 FoV. 
These values are in rough agreement with the values of $n = 4.09 \pm  0.48$ and $R\rm_e =$ 1258.46 $\pm$ 977.40 pc reported by \citet{Fabricius_2012}, which were obtained by fitting one-dimensional surface brightness profile combining multiple data sources. We caution that the disk effective radius may be underestimated due to the limited WFC3 FoV. 
The relative contribution of different components within the PPAK FoV is given in Table \ref{tab1}, which shows that the bulge is dominant over the disk. 
\\

\begin{figure}
\centering
\includegraphics[width=6in]{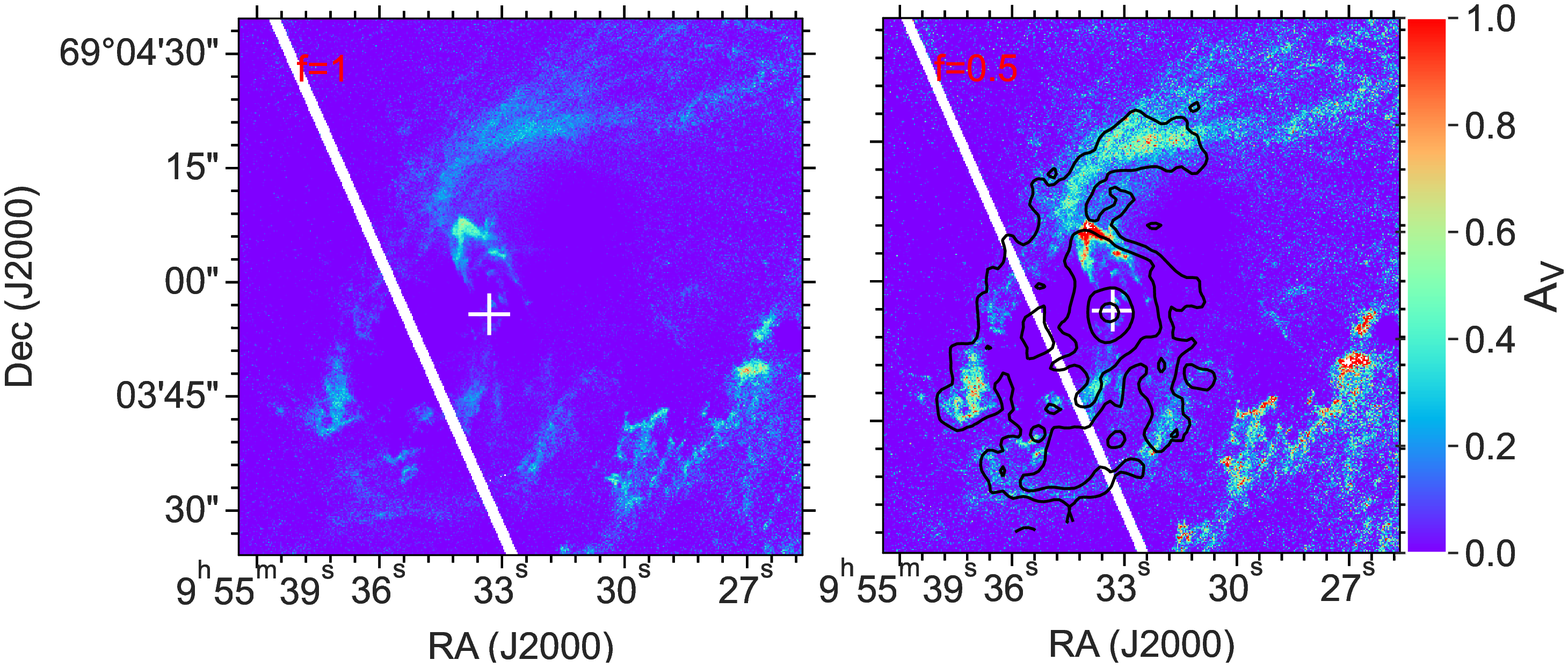}
\caption{$Left$: Absolute extinction derived from the WFC3/F438W residual image, for the $f=1$ case. $Right$: Similar to the left panel but for the $f=0.5$ case, overlaid with the HST H$\alpha$ intensity contours. 
The nucleus is marked by the cross. The white stripe is due to the chip gap of WFC3.
\label{fig:av}}
\end{figure}

Once the starlight model is subtracted from the WFC/F438W image, dust extinction features are readily revealed, which exhibit the familiar morphology of the nuclear spiral (Figure \ref{fig:galfit}). 
The apparent dust extinction along a given line-of-sight can be evaluated as $A_{\rm \lambda,app} = -2.5 \times \rm log (F\rm_{\lambda,obs}/F\rm_{\lambda,int})$, where $F\rm_{\lambda,obs}$ is the observed flux (e.g., in the F438W band) and $F\rm_{\lambda,int}$ is the intrinsic flux represented by the above model. 
Assuming a thin screen geometry, which is reasonable as suggested by the spatial coincidence between the dust extinction features and the nuclear spiral,
only the bulge starlight from behind this screen would be attenuated.
To account for this effect, we follow \cite{2016MNRAS.459.2262D} to introduce a parameter ($f$), i.e., the fraction of obscured starlight along the line-of-sight, such that $F\rm_{\lambda,obs}=\it [f\times \rm 10^{-0.4 \times \it A_{\rm \lambda,abs}}+(1-\it f)]\times F\rm_{\lambda,int}$, where $A_{\rm \lambda,abs}$ is the absolute extinction. 
Here we consider two special cases: 1) $f=1$, which occurs when the dusty screen is located in front of the bulge, and 2) $f = 0.5$, which occurs when the dusty screen is embedded in the bulge and seen face-on. 
The latter case might be closer to reality. 
We thus calculate the absolute extinction in the F438W band (central wavelength 4325$\rm~\AA$) and convert it into the more conventional V-band (5470$\rm~\AA$) extinction $A\rm_V$, assuming the Galactic extinction law of \cite{1989ApJ...345..245C}. 
The resultant V-band extinction maps for the PPAK FoV are shown in left and right panels of Figure \ref{fig:av}, for the case of $f = 1$ and $f = 0.5$, respectively. 
The $f = 0.5$ case exhibits systematically higher values than the $f = 1$ case, due to the negative correlation between $f$ and $A_{\rm \lambda,abs}$ for a given $A_{\rm \lambda,app}$.
The hydrogen column density $N\rm_H$ can thus be inferred to range between $\sim10^{20}~\rm cm^{-2}$ to a few times 10$^{21}~\rm cm^{-2}$ in the $f=1$ case, using the relation $N\rm_H \simeq 2 \times 10^{21} \it A\rm_V$ \citep{2009MNRAS.400.2050G}.
The $f=0.5$ case results in higher $N\rm_H$, reaching $10^{22} ~\rm cm^{-2}$ in some regions. 
\\

In Figure~\ref{fig:zoom-in}, the central $30\arcsec \times 30\arcsec$ is highlighted in H$\alpha$ emission and dust extinction. Under this zoom-in view, the bright core is seen to be connected with arm-like features to the southeast and southwest, as indicated by the red arrows in the figure, which have apparent counterparts in the dust extinction map. These arms might be feeding gas to the galactic nucleus.
Strong dust extinction features (indicated by red color) are also seen to the north of the nucleus, possibly with co-spatial H$\alpha$ emission that connects with the bright core.

\begin{deluxetable*}{cccccc}
\tablecaption{{\sc galfit} results}
\tablenum{1}
\tablewidth{2pt}
\tablehead{
\colhead{Component} & \colhead{P.A.\tablenotemark{(a)}} & \colhead{$R_{\rm e}$\tablenotemark{(b)}} & \colhead{Axis ratio\tablenotemark{(c)}} & \colhead{S$\rm \acute{e}$rsic index} & \colhead{Relative flux\tablenotemark{(d)}}\\
& \colhead{(degree)} & \colhead{(pc)} &  & 
}
\startdata
Bulge & 150.28 $\pm$ 0.02 & 999 $\pm$ 5 & 0.70 & 3.23 $\pm$ 0.01 & 1.0 \\
Disk & 149.67 $\pm$ 0.03 & 2472 $\pm$ 5 & 0.60 & 1 & 0.2\\
Nucleus & - & - & - & - & 0.002\\
\enddata
\tablecomments{(a) Position angle of the fitted component. (b) Effective radius. (c) Projected axis ratio. (d) Flux relative to that of the bulge component, integrated over the PPAK FoV. \label{tab1}}
\end{deluxetable*}

\subsection{Kinematics} \label{sec:3.2}

\subsubsection{Broad lines}
\label{subsubsec:broad}
\begin{figure}
\centering
\includegraphics[width= 6in]{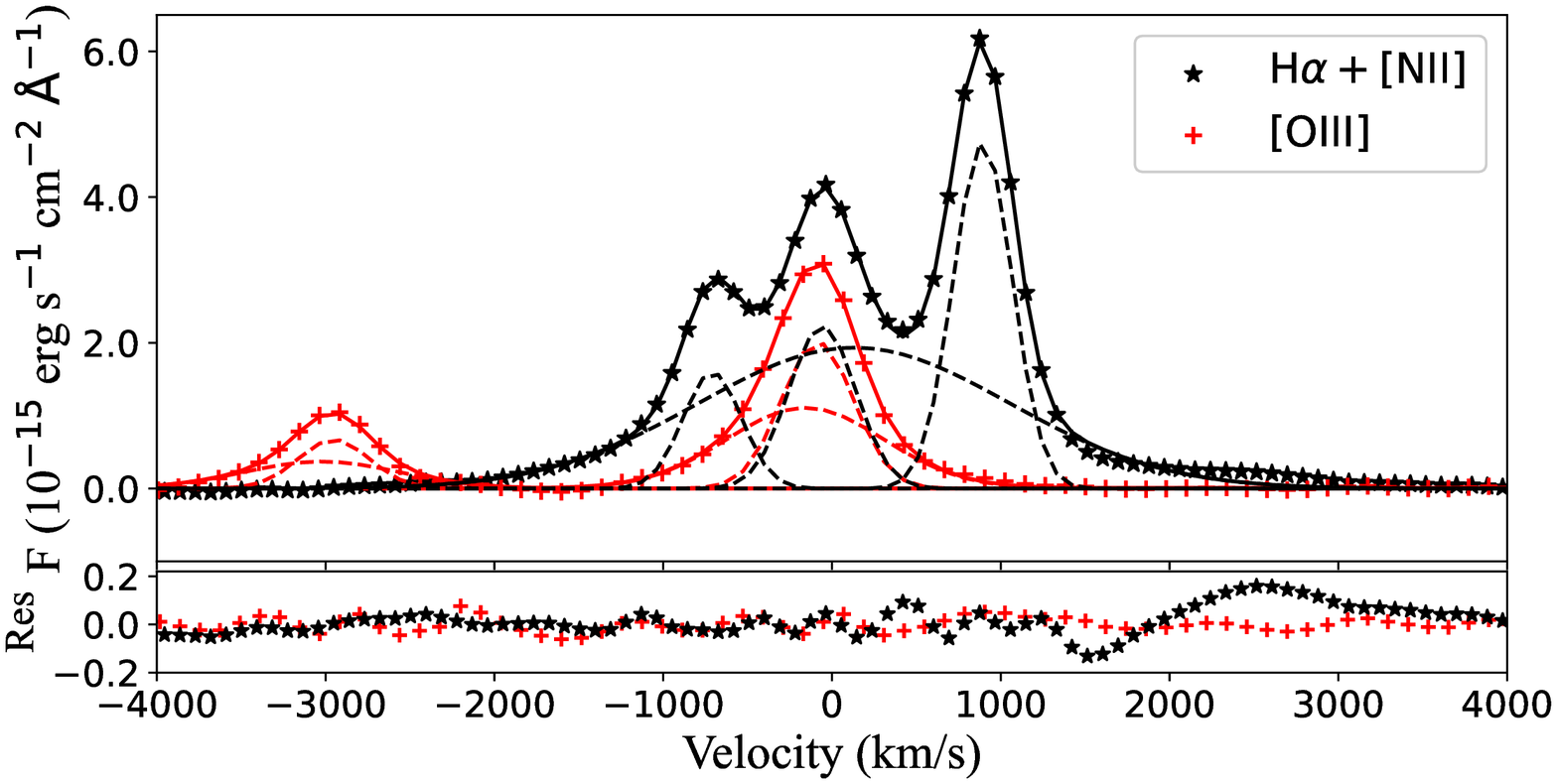}
\includegraphics[width= 3in]{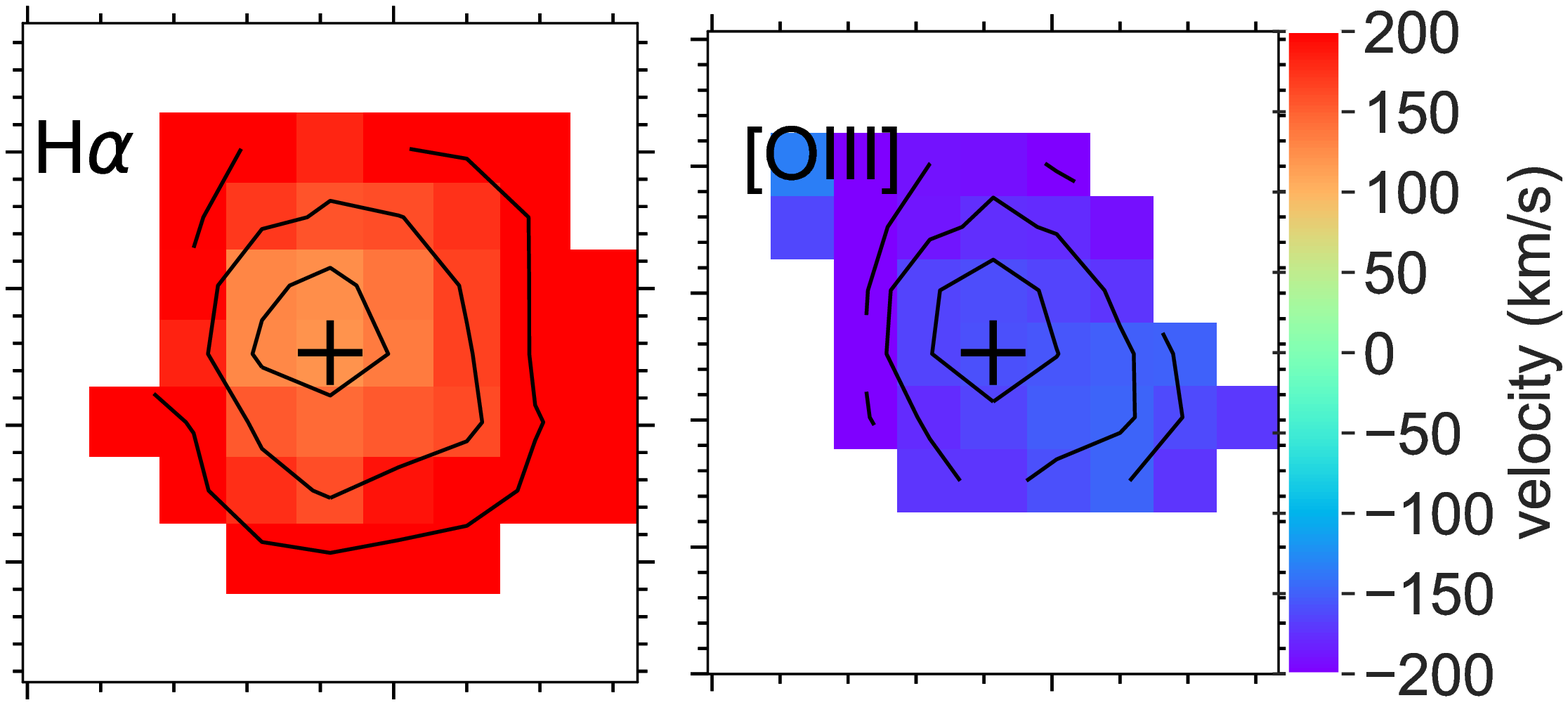}
\includegraphics[width= 3in]{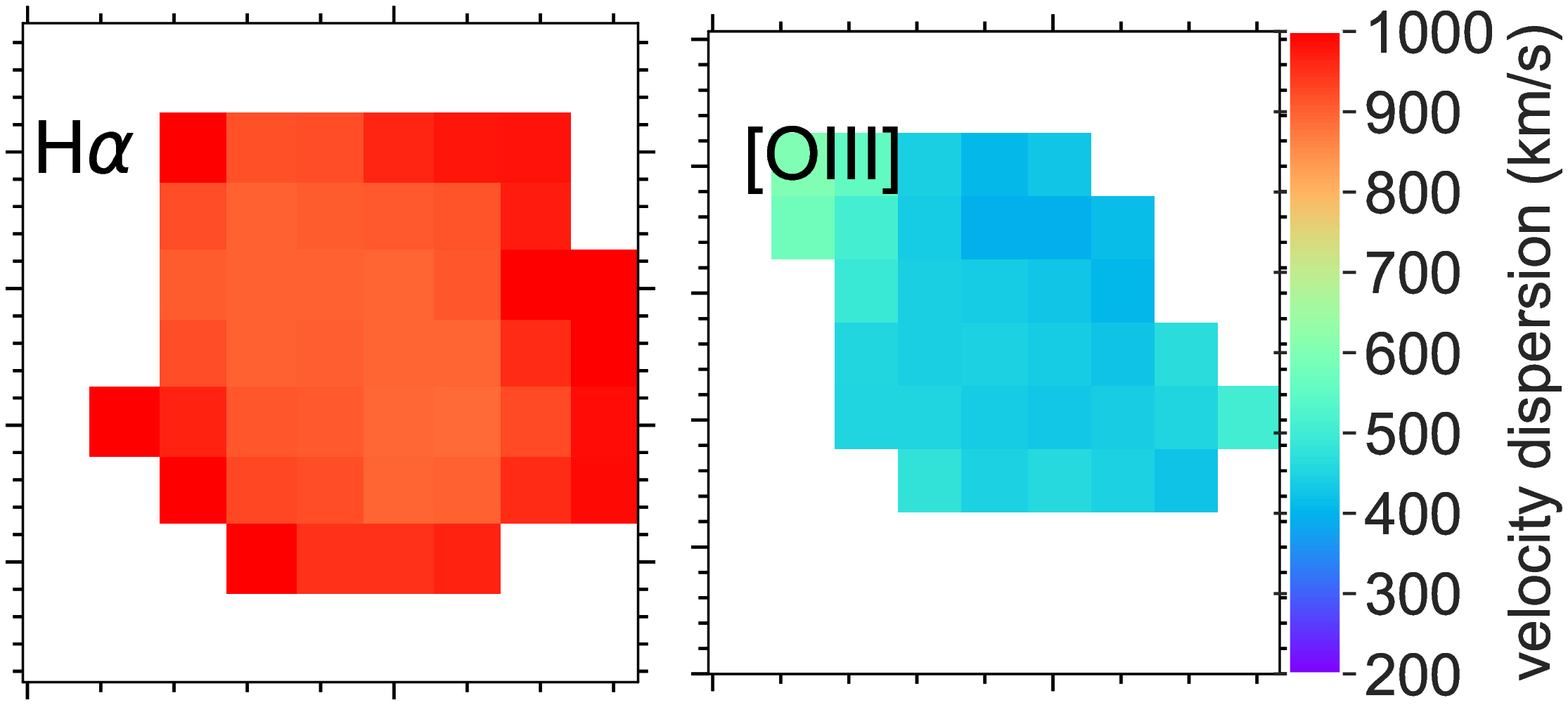}
\caption{$Upper$: Spectra of H$\alpha$ (black asterisks) and [O\,{\sc iii}] (red crosses) of the innermost pixel. The broad and narrow components of H$\alpha$ and the [N\,{\sc ii}] doublet are fitted with four Gaussians (black dashed lines), while the [O\,{\sc iii}] doublet are each fitted with a broad plus a narrow component (red dashed lines). The total multi-Gaussian fit is shown by the solid line, and the residual is plotted. $Lower$: Line-of-sight velocity (left) and velocity dispersion (right) maps of the broad component of H$\alpha$ and [O\,{\sc iii}] in the central $9\arcsec \times 9\arcsec$, overlaid with intensity contours with levels 1, 3, 7$\times 10^{-14}$ erg s$^{-1}$ cm$^{-2}$ arcsec$^{-2}$ for H$\alpha$, and 3, 7, 15$\times 10^{-15}$ erg~s$^{-1}$ cm$^{-2}$ arcsec$^{-2}$ for [O\,{\sc iii}]. The black cross marks the central pixel (1$\arcsec$ size). 
\label{fig:spec}}
\end{figure}

As demonstrated in previous work \cite[e.g.,][]{1996AJ....111.1901B, 1996ApJ...462..183H}, the nucleus of M81 exhibits a broad (FWHM $\gtrsim$ 1000 km~s$^{-1}$) component in the H$\alpha$ line, which is a typical signature of the broad-line region of an AGN. The PPAK data allows us to probe the existence of a broad line at regions beyond the nucleus.
To do so, we fit the H$\alpha$ line of the central 9$\arcsec$ $\times$ 9$\arcsec$ region with a double-Gaussian profile, one for the narrow component and the other for the putative broad component. 
In the upper panel of Figure \ref{fig:spec}, we show the fitted spectrum in the interval around H$\alpha$ of the innermost pixel as an illustration (black symbols and curves). 
Two additional Gaussians are included to account for the [N\,{\sc ii}] doublet, 
as well as a power-law to account for the baseline continuum.
In this central pixel, the velocity dispersion of H$\alpha$ is $\sigma_{\rm H\alpha} = 913 \pm 3$ km s$^{-1}$ for the broad component, and $\sim 100$ km s$^{-1}$ for the narrow component (comparable to the instrumental broadening, which has been subtracted). Errors have been estimated from 200 iterations of Monte Carlo simulations. 
This result is in good agreement with the value of 862 $\pm$ 37 km s$^{-1}$ obtained from the GMOS spectrum of the central 1.5$\arcsec$ \citep{2011MNRAS.413..149S}. 
It is found that a broad H$\alpha$ component is prevalent in this central region, with a radial extent of at least 3$\arcsec$ ($\sim$50 pc) in all directions, as illustrated in the lower panel of Figure \ref{fig:spec}, where the spatial distribution of the best-fit line-of-sight velocity and velocity dispersion of the broad component are plotted.
We note that a broad component is not required to provide a satisfactory fit to the [N\,{\sc ii}] doublet.
\\

On the other hand, we find that the [O\,{\sc iii}] line in this region also exhibits the signature of a broad component, as illustrated in the upper panel of Figure \ref{fig:spec} (red symbols and curves), again using the innermost pixel as an example. 
We thus fit the [O\,{\sc iii}] line with a double-Gaussian profile. 
The resultant broad component of [O\,{\sc iii}] has a velocity dispersion of $434$ km~s$^{-1}$, which is about half that of the broad H$\alpha$, but is significantly higher than that of the narrow component ($\sim 100$ km~s$^{-1}$).
As evident in the lower panel of Figure \ref{fig:spec}, the broad component of [O\,{\sc iii}] shares the broad H$\alpha$ with a similar spatial extent and a trend of radially decreasing intensity, which suggest a common origin between the two lines.
\\

However, the broad H$\alpha$ and [O\,{\sc iii}] components are distinct in their kinematics. As seen in Figure \ref{fig:spec}, the H$\alpha$ is collectively redshifted, with a mean line-of-sight velocity $v_{\rm H\alpha} = 197 \pm 5$ km s$^{-1}$, and has a mean velocity dispersion of $\sigma_{\rm H\alpha} = 940 \pm 14$ km s$^{-1}$. 
In contrast, the [O\,{\sc iii}] is collectively blueshifted, with $v_{\rm [O\,{\sc III}]} = -176 \pm 5$ km s$^{-1}$, and has a substantially lower velocity dispersion of $\sigma_{\rm [O\,{\sc III}]} = 454 \pm 9$ km s$^{-1}$. 
Here the line-of-sight velocity has not been corrected for the systemic velocity of M81 (-39 km~s$^{-1}$, \citealp{de_Blok_2008}).
A redshifted broad H$\alpha$ line was also reported based on HST/STIS spectra \citep{2003AJ....125.1226D, 2007ApJ...671..118D, 2019MNRAS.488.1199D}, with $v = 192$ km s$^{-1}$ and $\sigma = 930 \pm 11$ km s$^{-1}$. Interestingly, from the STIS spectrum covering the central $0.2\arcsec\times 0.35\arcsec$, \cite{2019MNRAS.488.1199D} reported a smaller velocity dispersion ($\sigma \approx 250$ km s$^{-1}$) for the [O\,{\sc iii}]$\lambda$5007 line, which is intermediate between the values of the broad and narrow components found here. 
 Unfortunately, the [O\,{\sc iii}] line was beyond the wavelength coverage of the GMOS IFU observation \citep{2011MNRAS.413..149S}. \\

The apparent radial extent of the broad H$\alpha$ and [O\,{\sc iii}] lines is substantially larger than previously known, suggesting the presence of a high-velocity and high-excitation plasma within the central 50 pc of M81. 
As a cautionary note, the effect of PSF scattering (FWHM $\lesssim 2.4\arcsec$; \citealp{2015A&A...576A.135G}) of the bright nucleus could have led to an artificially large broad-line region. 
Future IFU observations with a higher angular resolution are needed to confirm this case.
\\

\subsubsection{Narrow lines}
\label{subsubsec:narrow}
\begin{figure}
\centering
\includegraphics[width= 7in]{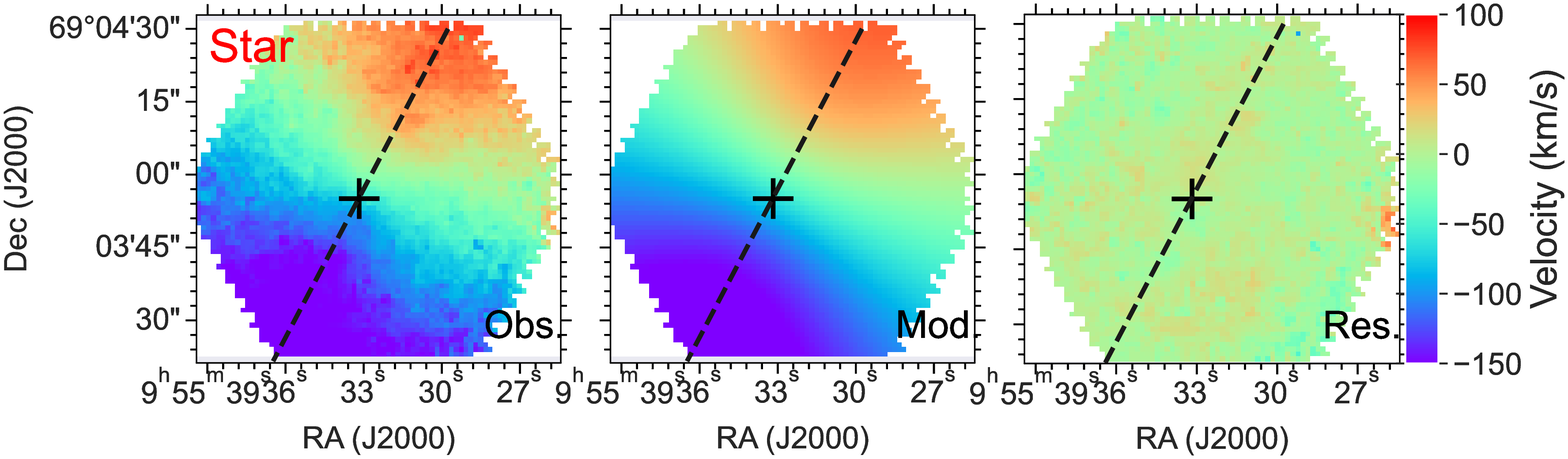}
\includegraphics[width= 7in]{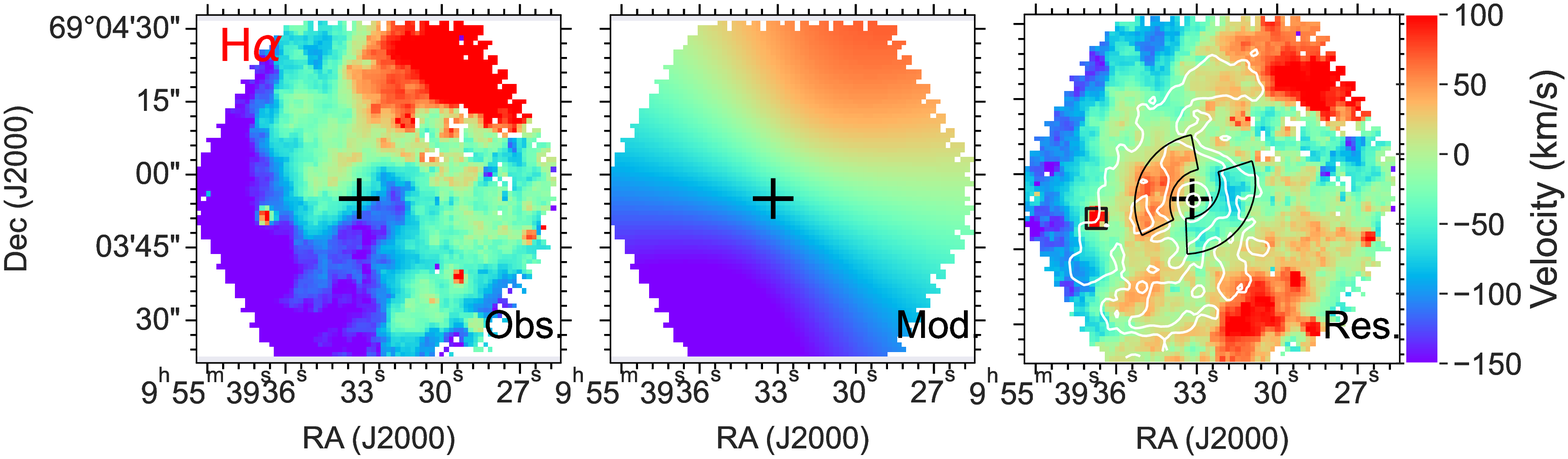}
\includegraphics[width= 7in]{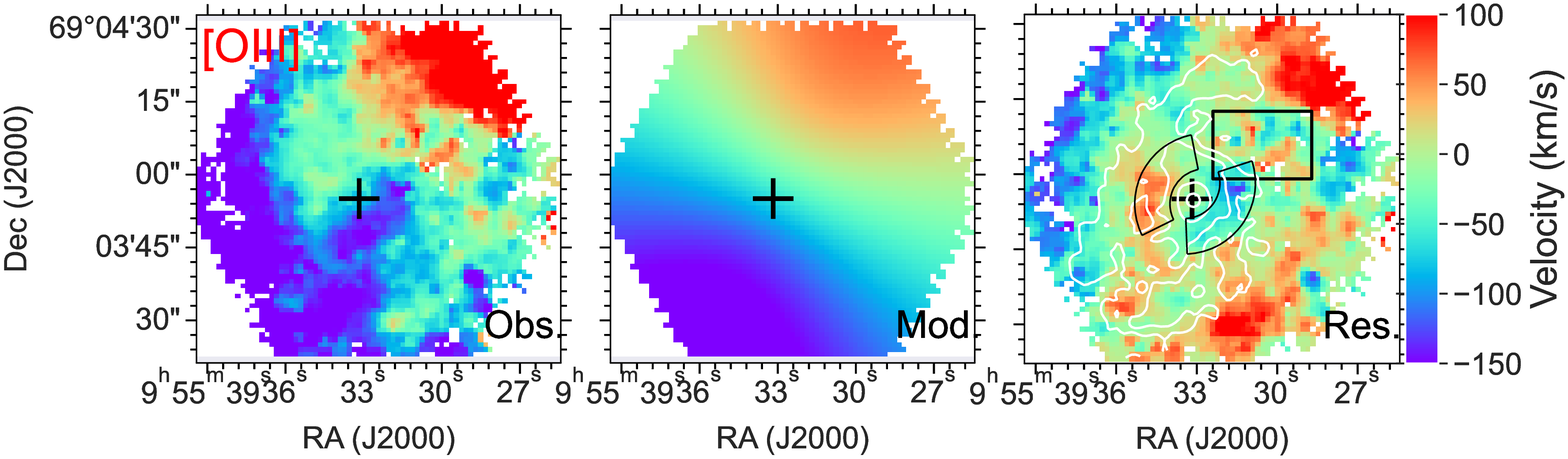}
\caption{The velocity field of stars, H$\alpha$ and [O\,{\sc iii}], from top to bottom. 
$Left$: The observed two-dimensional distribution of line-of-sight velocity. $Middle$: the best-fit kinematic model of the stellar velocity field. The same model is adopted as the base model for H$\alpha$ and [O\,{\sc iii}]. 
$Right$: the residual velocity map by subtracting the model from the observed velocity field. The dashed line marks the kinematic major-axis, and the cross marks the galactic nucleus. In the middle and bottom right panels, the two fan-shaped regions outline a putative bi-conical outflow, and the HST H$\alpha$ intensity contours are overlaid. The small and large black boxes in H$\alpha$ and [O\,{\sc iii}] maps, respectively, outline regions with a significant redshifted component as shown in Figure \ref{fig:dv}. \label{fig:v}}
\end{figure}

We now turn to the narrow lines that are prevalent throughout almost the entire PPAK FoV, focusing on the kinematics of the H$\alpha$ and [O\,{\sc iii}] lines.
The observed line-of-sight velocity of H$\alpha$ and [O\,{\sc iii}] is shown in the left panels of Figure \ref{fig:v}, along with the stellar line-of-light velocity derived from the continuum model.
Overall, the stars show a regular and smooth velocity field with an obvious rotation pattern, while the H$\alpha$ and [O\,{\sc iii}] lines follow this rotation pattern but exhibit a more complex velocity field. \\

We apply a simple analytical model of \citet{1991ApJ...373..369B}, which assumes a circular orbit, to fit the line-of-sight velocity of the stars, as a function of the projected radius ($R$) and position angle ($\Psi$, east from north),
\begin{equation}
{\label{eqn:v}}
V\rm_{mod}\it (R, \rm \Psi) = \it V_s + \frac{AR\rm cos(\Psi - \Psi_0)sin(\it i)\rm cos\it^p(i)}{\{R\rm^2[\rm sin^2(\Psi - \Psi_0) + cos^2(\it i)\rm cos^2(\Psi - \Psi_0)] + \it C\rm_0^2 cos^2(\it i)\}^{p/\rm 2}},
\end{equation}
where $V\rm_s$ is the systemic velocity, $\rm \Psi_0$ is the position angle of the kinematic major-axis, $i$ is the inclination angle of the orbital plane, $A$ is the maximum orbital velocity, $C\rm_0$ is the radius at the maximum velocity, and $p$ is the slope of the rotation curve outside of $C\rm_0$. 
Due to the moderate FoV, the parameters $A$, $C\rm_0$ and $p$ are degenerate. Fortunately, this degeneracy does not have a significant effect on the other parameters, hence we will not include them in the following discussion. 
\\

The best-fit parameters are derived using MCMC simulations, which result in $V\rm_s = -56.60^{+0.18}_{-0.24}$ km~s$^{-1}$, $i = 34.^\circ 8^{+1.2}_{-1.2}$, and $\rm \Psi_0 = 153.^\circ 22^{+0.14}_{-0.15}$.
The systemic velocity is somewhat more blueshifted than the canonical value of $-39 \pm 3$ km s$^{-1}$, which was obtained from the HI kinematics primarily tracing the outer disk of M81 \citep{de_Blok_2008}. 
The inclination angle is also significantly lower than the value of 58$^\circ$ estimated from the axis ratio of the M81 disk\footnote{NASA/IPAC Extragalactic Database}. 
This may be understood as the velocity field in this region being dominated by the bulge rather than the disk (Table \ref{tab1}). 
Nevertheless, the excellent agreement between the fitted position angle of the kinematic major-axis (marked by the dash line in the upper panels of Figure \ref{fig:v}) and that of the disk (153$^\circ$, \citealp{1991rc3..book.....D}) suggests that we have arrived at a reasonable phenomenological model of the stellar velocity. 
This is further supported by the the difference map between the observed and modelled stellar velocity field (the upper right panel of Figure \ref{fig:v}), which shows no significant residual patterns.
From the optical image of M81 in Figure \ref{fig:fov}, which exhibits more prominent extinction features on the southwestern side of the disk, we can conclude that this side is the near side. Accordingly, it can be inferred that the disk rotates counterclockwise, which is consistent with a trailing spiral arm pattern. 
\\

We then examine potential kinematic signatures of the ionized gas beyond the coherent rotation. 
For this purpose we subtract the above stellar kinematic model from the observed gas velocity field. This implicitly assumes that to zeroth order the bulk of the ionized gas has the same underlying rotation pattern as that of the stars. 
This is verified by fitting an independent model of Eqn.~\ref{eqn:v} to the H$\alpha$ and [O\,{\sc iii}] velocity fields, both of which result in best-fit parameters that are in rough agreement with the stellar velocity field.
Nevertheless, we take the latter as the underlying base model because it is free of complex kinematic features.  
The residual maps are shown in the right panels of Figure \ref{fig:v}, which signify the non-rotating components of the velocity field.
It is evident that the residual velocity field is highly similar between the H$\alpha$ and [O\,{\sc iii}] lines, perhaps not surprising because the same base velocity field has been subtracted.
\\

Perhaps the most interesting pattern in the residual velocity map is a pair of redshifted/blueshifted arc-shaped features roughly symmetric about the nucleus, as outlined in the middle right and lower right panels of Figure \ref{fig:v}.
Both features are located at a projected radius of $\sim$120--250 pc from the nucleus and have an absolute line-of-sight velocity of $\sim$ 50 km s$^{-1}$ with respect to the nucleus.
A plausible explanation for these ``arcs'' is that they are the manifestation of a bi-conical outflow from the nucleus, which lies predominantly in or near the disk plane, such that the blueshifted component is found at the southwestern side (near side) and the redshifted component is found at the northeastern side (far side). 
It is noteworthy that these two ``arcs'' are not part of the nuclear spiral, rather they fall in between the core and the nuclear spiral (Figure~\ref{fig:zoom-in}).
Alternatively, a bi-polar inflow along the disk rotation axis would produce similar velocity patterns, which, however, appears physically implausible. 
Non-circular motions caused by a nuclear bar or triaxial bulge could be another possibility, but the observed velocity appears too high to be compatible with typical pattern speed of bars \citep[$20-50\rm~km~s^{-1}~kpc^{-1}$;][]{2015ApJ...806..150L}.
Support for an outflow comes from \cite{2011MNRAS.413..149S} and \cite{2015A&A...576A..58R}, who both analyzed the velocity field in the central 10 pc using the principle component analysis (PCA) tomography \citep{2009MNRAS.395...64S}, and interpreted one of the components (PC 4) as a conical outflow of ionized gas. 
Also prominent in the residual velocity map are redshifted features at the southwestern and northwestern corners of the FoV (appearing in red color), as well as blueshifted features at the northeastern edge.  
Judging from the dust extinction map (Figure~\ref{fig:av}), the redshifted features might be part of the outer nuclear spiral. These features may signify inflowing motions along the disk plane. We caution that the line emission in these regions are relatively weak, which may be more vulnerable to systematic effects. Future IFU observations of larger FoV are warranted to test and better quantify the putative inflowing motion associated with these features. 
\\

\begin{figure}
\centering
\includegraphics[width= 3.5in]{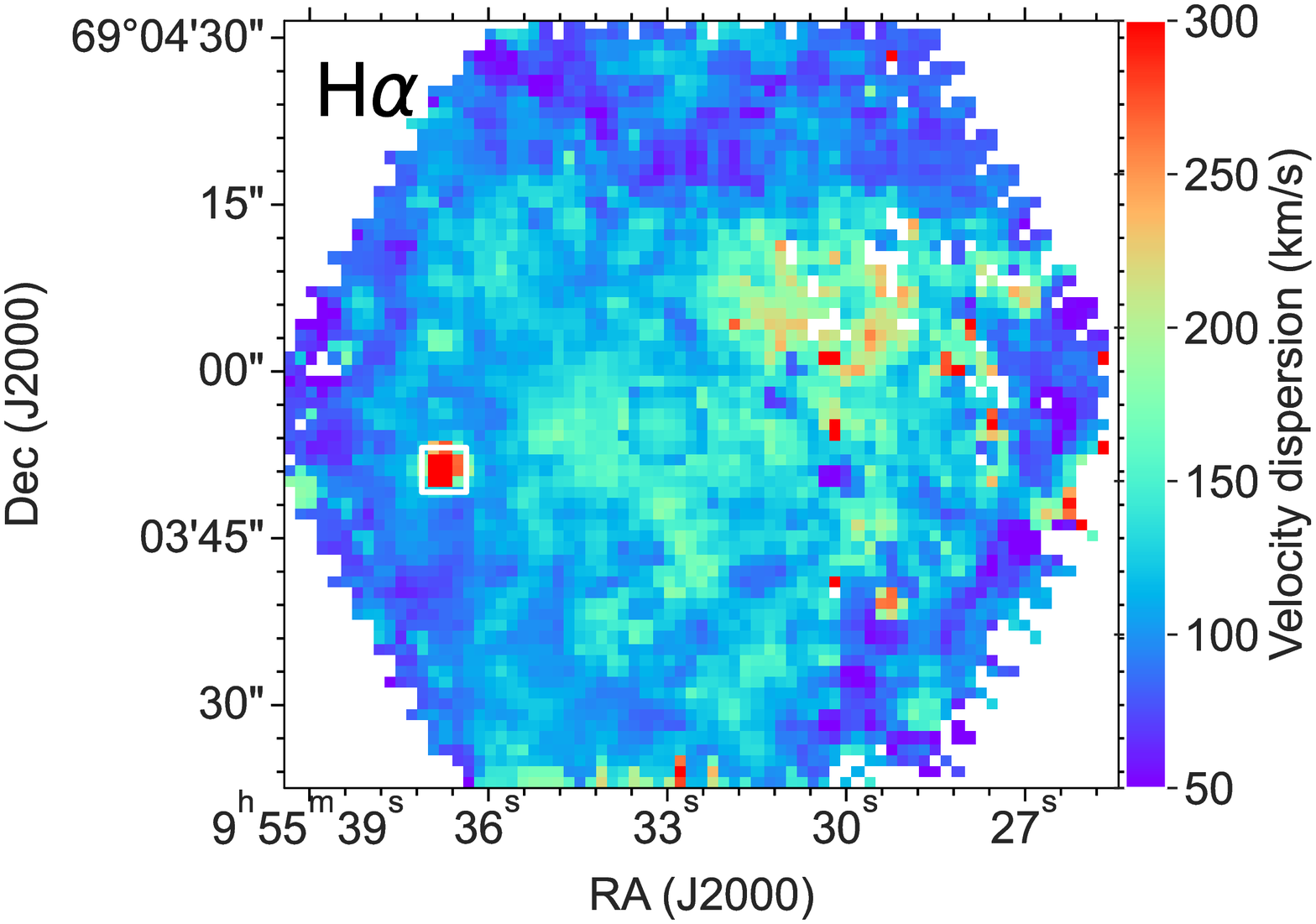}
\includegraphics[width= 3.5in]{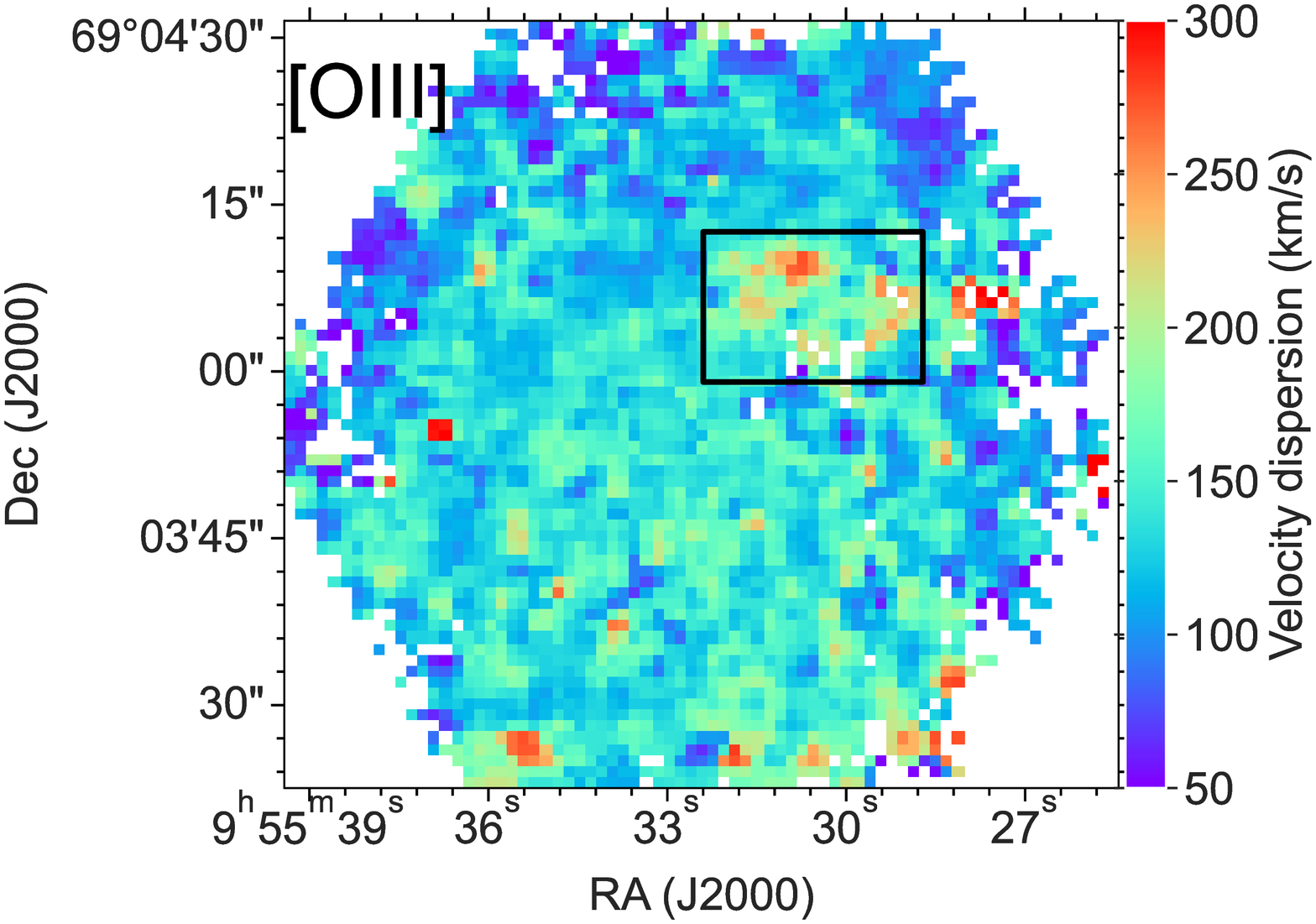}
\includegraphics[width=3.5in]{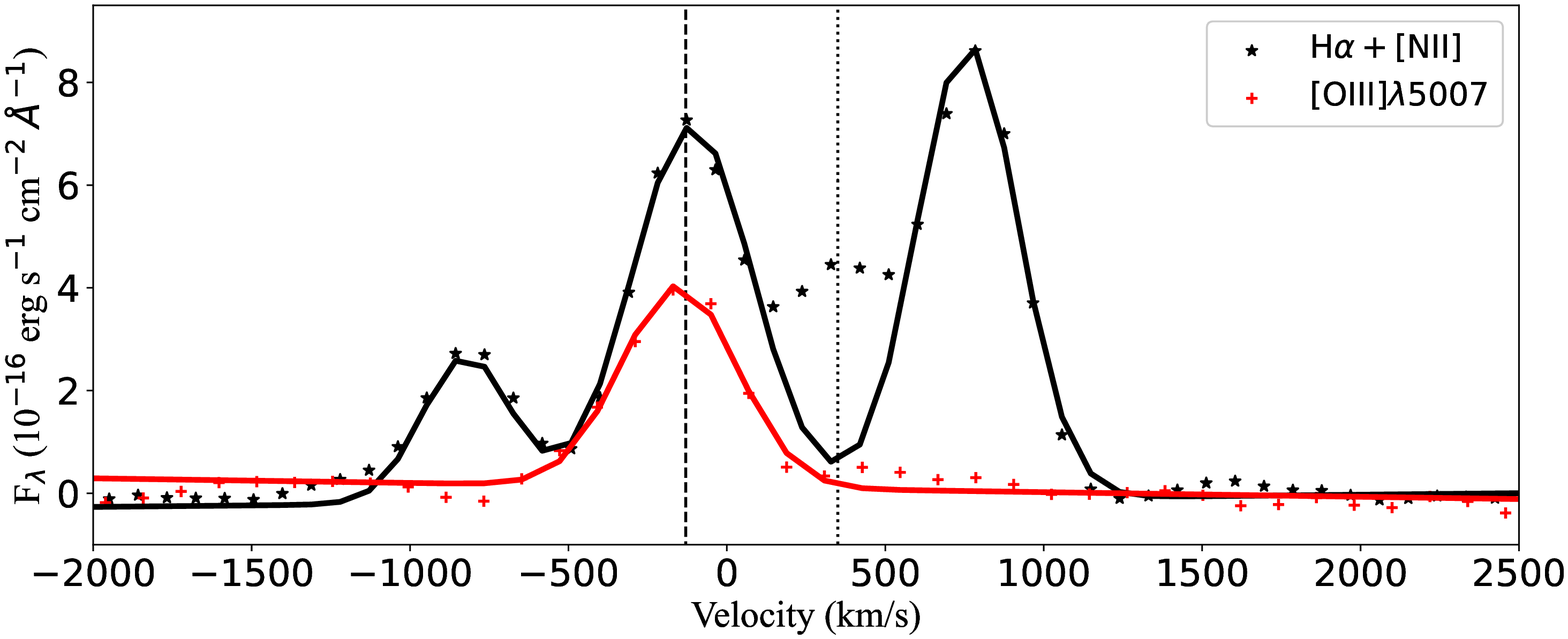}
\includegraphics[width=3.5in]{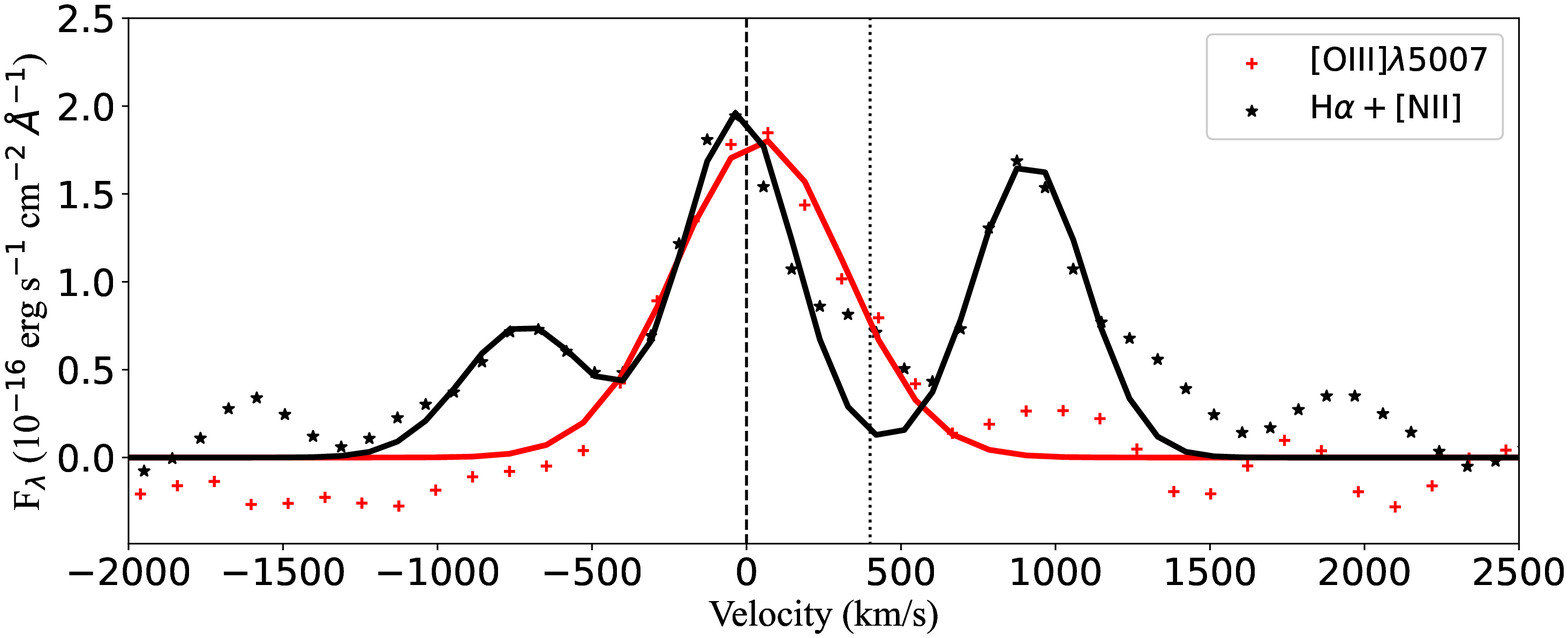}
\caption{$Upper~Left$: H$\alpha$ velocity dispersion. The white box delineates a clump with high H$\alpha$ velocity dispersion, which is also significantly redshifted with respect to its vicinity. 
$Upper~Right$: [O\,{\sc iii}] velocity dispersion. The black box shows a region with  high velocity dispersion in [O\,{\sc iii}]. $Lower~Left$: Stacked spectra of H$\alpha$ + [N\,{\sc ii}] (black) and [O\,{\sc iii}] (red) within the white box. The primary velocity component of H$\alpha$ is indicated by the vertical dashed line. Excess on the red wing of H$\alpha$ is evident and indicated by the dotted line. $Lower~Right$: Stacked spectra of [O\,{\sc iii}]$\lambda5007$ (red) 
and H$\alpha$ + [N\,{\sc ii}] (black) from pixels inside the black box with $\sigma\rm_{[O\,III]} > $ 250 km s$^{-1}$. The dashed line marks the primary velocity component. Possible excess at the red wing of H$\alpha$, in accordance with the broader [O\,{\sc iii}], is indicated by the dotted line. 
\label{fig:dv}}
\end{figure}

Figure \ref{fig:dv} displays the distribution of velocity dispersion of H$\alpha$ and [O\,{\sc iii}].
The two lines show overall smooth and similar distributions, with a comparable level of mean velocity dispersion ($\sim$120 km~s$^{-1}$; also comparable to the level of instrumental broadening, which has been subtracted). 
We note that typical uncertainty of the velocity dispersion is $\lesssim 5$ km s$^{-1}$ in the nuclear spiral, and somewhat larger ($\sim 15-20$ km s$^{-1}$) in the outer regions with lower S/N.
There is no significant feature in the central region after excluding the broad component. A region of high velocity dispersion, with values $\gtrsim$250 km s$^{-1}$, is seen in the northwestern side (outlined by the black rectangle) of the [O\,{\sc iii}] map.  
The stacked spectra of [O\,{\sc iii}]$\lambda$5007 and H$\alpha$+ [N\,{\sc ii}] of pixels having $\sigma\rm_{[O\,III]} > $ 250 km s$^{-1}$ in this region are presented in the lower right panel of Figure \ref{fig:dv}. 
Some excess is seen on the red wing of both H$\alpha$ and [N\,{\sc ii}]$\lambda$6584, possibly in accordance with the broad [O\,{\sc iii}]. The physical origin of this feature is unclear and deserves further exploration.
\\

A compact feature of high velocity dispersion is additionally seen at the eastern side of the H$\alpha$ map (marked by the small white box). This clump-like feature is also prominent in the H$\alpha$ velocity map (Figure~\ref{fig:v}), which is redshifted with respect to its vicinity. 
The stacked spectra of this region are shown in the lower left panel of Figure \ref{fig:dv}. It can be seen that there is a second velocity component redshifted by $\sim$ 500 km s$^{-1}$ from the primary component (at $\sim$ -130 km s$^{-1}$) of H$\alpha$, which is however not seen in the [O\,{\sc iii}] line. 
The location of this clump coincides with a molecular cloud identified by \cite{2007A&A...473..771C} with IRAM 30m CO observations. However, the CO cloud is blueshifted, same as the primary component of the H$\alpha$ and H$\beta$. Hence the redshifted secondary component cannot be attributed to the CO cloud. 
We speculate that this clump is a low-excitation cloud in the M81 group. 
\\


\subsection{Line ratios} \label{sec:3.3}

Physical mechanisms responsible for the gas ionization can in principle be distinguished using line ratio diagnostics, among which the most commonly used is the BPT diagram \citep{Baldwin_1981}. 
We construct a spatially-resolved BPT diagram for the circumnuclear region of M81, using the standard pair of [N\,{\sc ii}]/H$\alpha$ and [O\,{\sc iii}]/H$\beta$ line ratios. 
In the present case, the number of useful pixels for the BPT diagram is mostly constrained by the relatively weak H$\beta$ line.
Therefore, to maximize the useful information, we apply the Voronoi tessellation binning \citep{2003MNRAS.342..345C}
on the continuum map with a target $S/N$ of 7000. 
This helps to reliably measure lines from low-intensity area typically found at large radii, while preserving the best spatial resolution of high-intensity area at small radii.  
The binned H$\beta$ intensity map is produced by measuring the H$\beta$ line using a single Gaussian from individual Voronoi bins. 
Only bins with $S/N$ $>$ 3 of the line flux are shown in this map, while bins with $S/N$ $<$ 3 are left blank. This results in a fractional valid area of 90\%, compared to 46\% without binning (Figure \ref{fig:flux}). 
\\

\begin{figure}
\centering
\includegraphics[width= 3.55in]{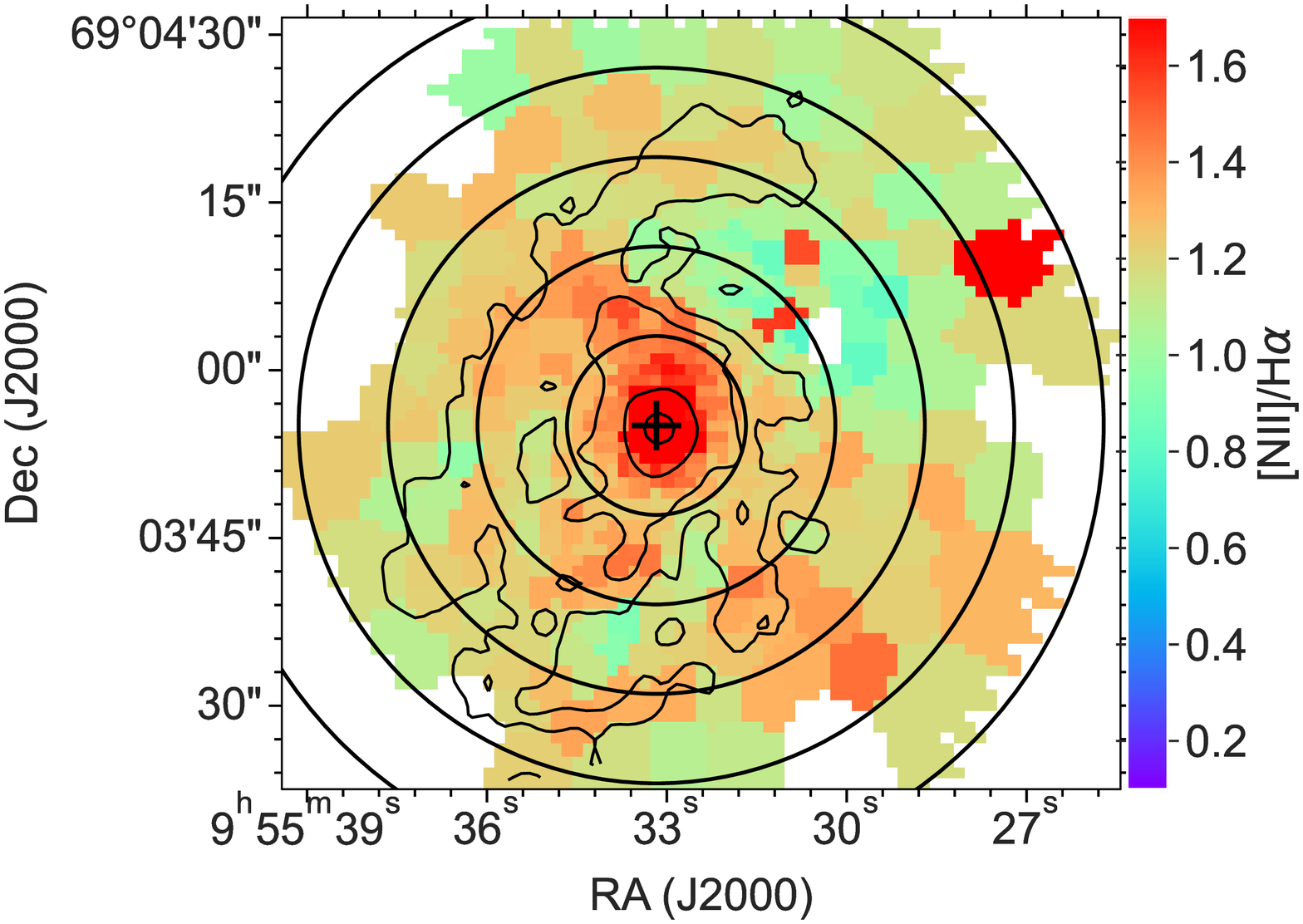}
\includegraphics[width= 3.43in]{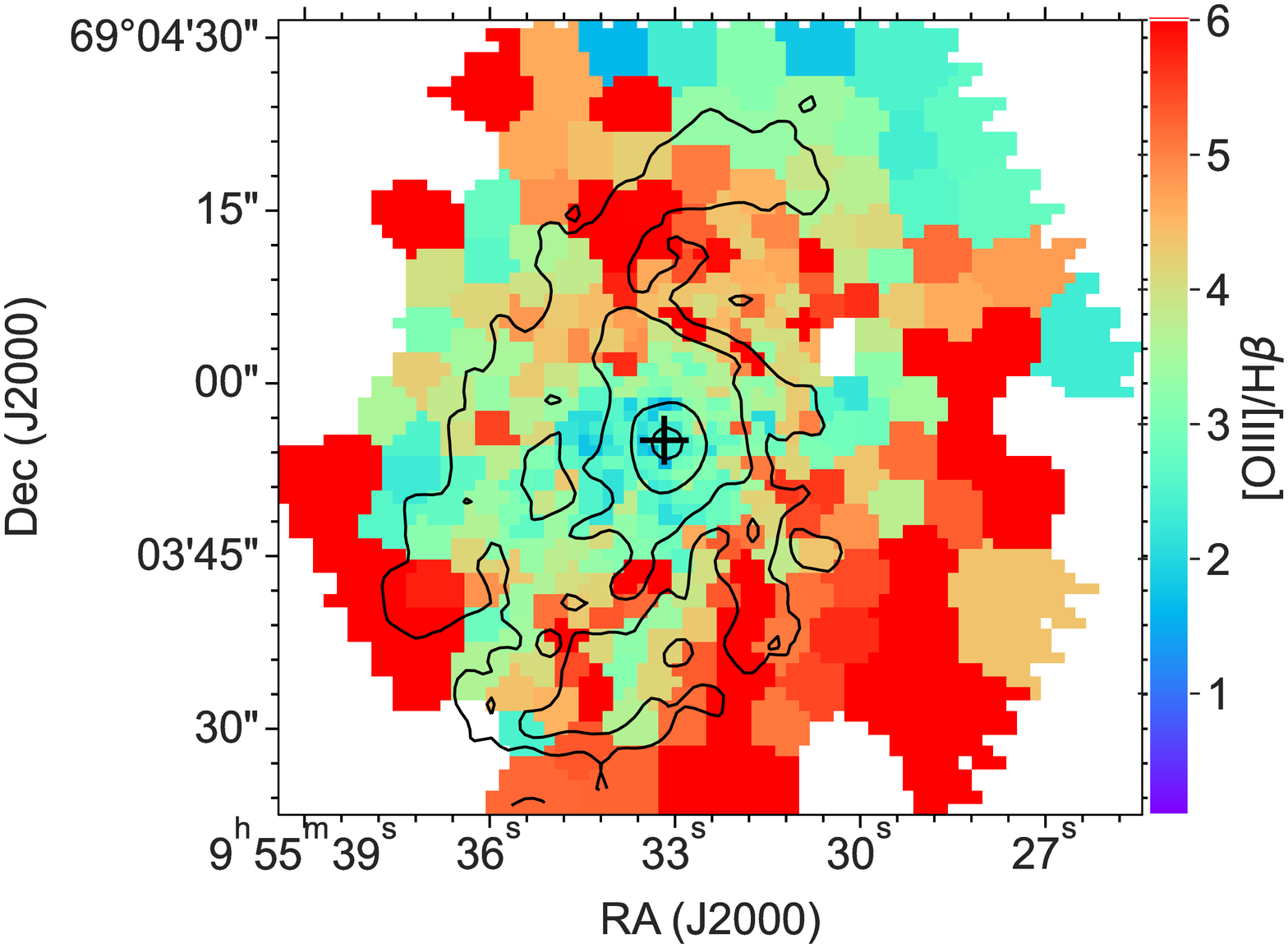}
\caption{The [N\,{\sc ii}]/H$\alpha$ (left) and [O\,{\sc iii}]/H$\beta$ (right) ratio maps after Voronoi tessellation binning, overlaid with the HST H$\alpha$ intensity contours. 
The ratios are plotted in a linear scale. The circles define the five annuli used in Figure \ref{fig:ratio}. 
In both panels, bins with  the H$\beta$ having $S/N$ $<$ 3 are left blank. The black cross marks the nucleus.
\label{fig:r2}}
\end{figure}

The binned [N\,{\sc ii}]/H$\alpha$ and [O\,{\sc iii}]/H$\beta$ flux ratio maps are shown in Figure \ref{fig:r2}. Here we have excluded the broad line component found in the central few pixels (Section~\ref{subsubsec:broad}). It can be seen that the [N\,{\sc ii}]/H$\alpha$ ratio is higher in the bright core region, and generally decreases outward. 
On the other hand, the [O\,{\sc iii}]/H$\beta$ ratio is lower in the core but increases in the nuclear spiral, as illustrated by the HST H$\alpha$ intensity contours. 
High values of [O\,{\sc iii}]/H$\beta$ found near the southwestern edge of the map is primarily driven by the weak H$\beta$ emission there. 
\\

\begin{figure}
\centering
\includegraphics[width= 3.5in]{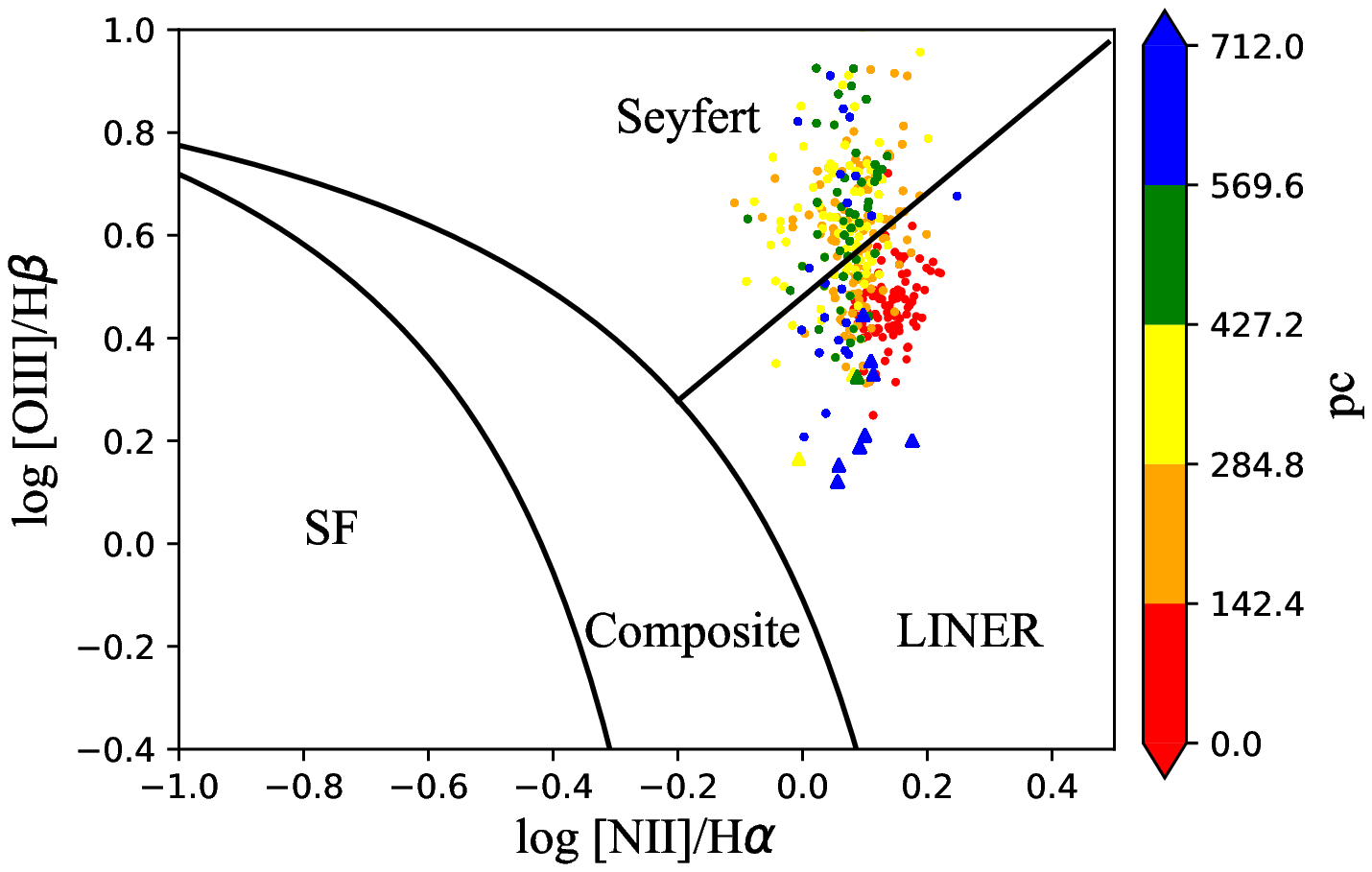}
\includegraphics[width= 3.5in]{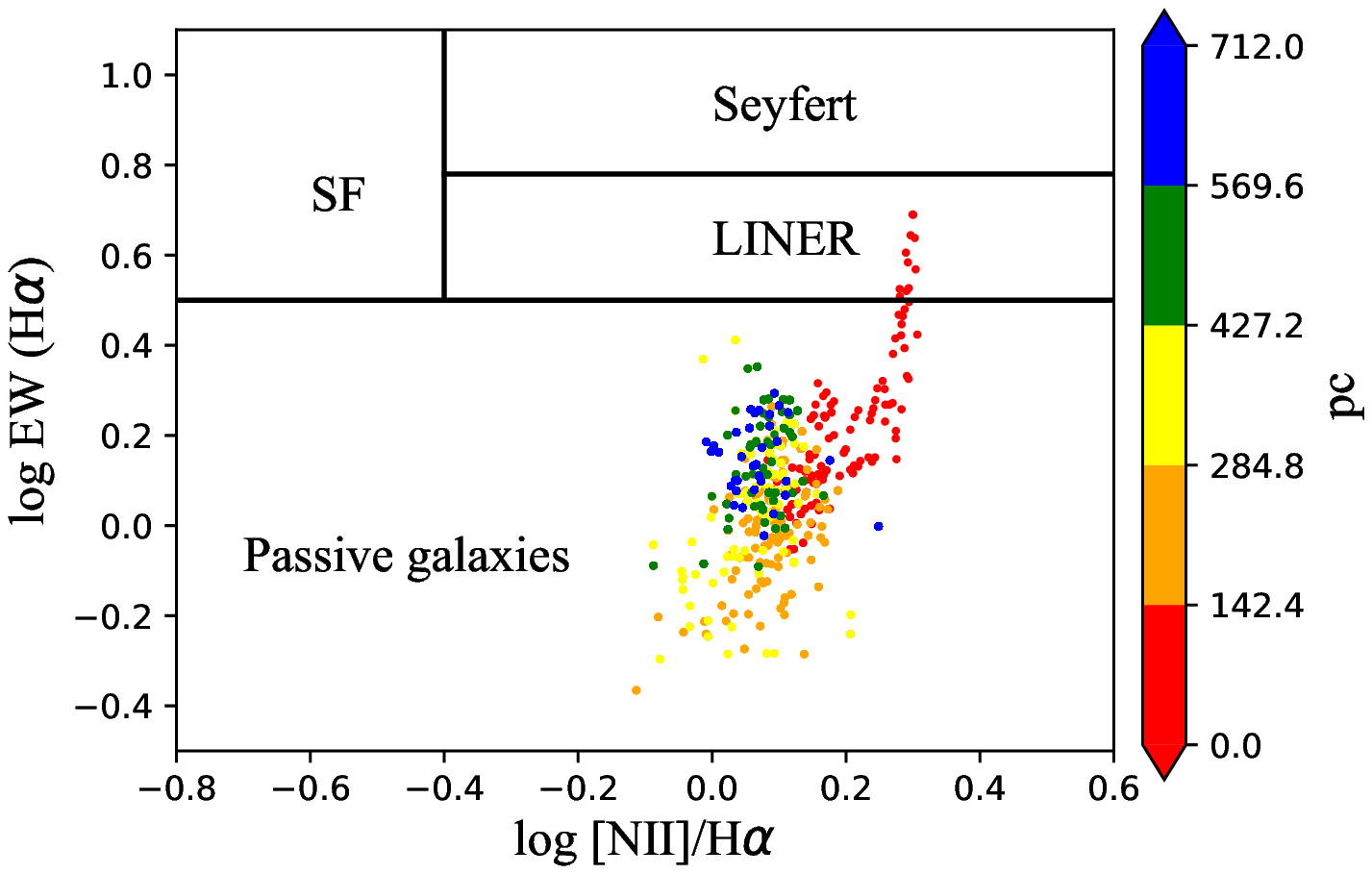}
\caption{$Left$: BPT diagram of [N\,{\sc ii}]/H$\alpha$ vs. [O\,{\sc iii}]/H$\beta$, consisting of all Voronoi bins and color-coded in terms of five radial ranges. The triangles represent 3$\sigma$ lower limits of the O\,{\sc iii}]/H$\beta$ ratio for bins with H$\beta$ $S/N$ $<$ 3. The solid lines are are from \cite{2001ApJ...556..121K} and \cite{2003MNRAS.346.1055K}, which divide the diagram into regimes of star formation, composite, Seyfert and LINER. 
$Right$: The WHAN diagram, showing [N\,{\sc ii}]/H$\alpha$ vs. the H$\alpha$ equivalent width for all Voronoi bins, color-coded in the same way as in the left panel. 
\label{fig:ratio}}
\end{figure}

The standard BPT diagram is produced using the [N\,{\sc ii}]/H$\alpha$ and [O\,{\sc iii}]/H$\beta$ ratios in individual Voronoi bins, as shown in the left panel of Figure \ref{fig:ratio}. The separating lines from \cite{2001ApJ...556..121K} and \cite{2003MNRAS.346.1055K} divide the diagram into characteristic  regimes of star formation (SF), Composite (SF + AGN), Seyfert, and LINER. 
The bins are color-coded in terms of their projected distance from the nucleus, grouped in five annuli. 
From this diagram, it is clear that the majority of bins, in particular those inside the core, can be classified as LINER. 
The remaining bins lie in the regime of Seyfert, whereas no single bin falls in the SF or composite regime. 
We also include Voronoi bins with H$\beta$ $S/N$ $<$ 3 in the diagram (triangles), which stand for a 3$\sigma$ lower limit of the [O\,{\sc iii}]/H$\beta$ ratio. 
Most such bins are found in the two outermost annuli and fall in the regime of LINER.
\\

In the right panel of Figure~\ref{fig:ratio}, we plot the so-called WHAN diagram \citep{2011MNRAS.413.1687C}, which contrasts the equivalent width of H$\alpha$ and the [N\,{\sc ii}]/H$\alpha$ ratio. 
While the WHAN diagram was originally proposed for a classification of SDSS emission-line galaxies, in particular those having weak lines (i.e., passive galaxies), it is relevant here since the circumnuclear region of M81 is relatively gas-poor (see discussions in Section~\ref{sec:4}) and the weak H$\beta$ in the outer annuli could be the main cause of a high [O\,{\sc iii}]/H$\beta$ ratio and consequently the appearance of Seyfert signature in the BPT diagram. In this WHAN diagram, indeed the majority of bins fall in the regime of passive galaxies, with only a few bins belonging to the core classified as LINER and no single bin as Seyfert or SF.
\\


\begin{figure}
\centering
\includegraphics[width= 3.3in]{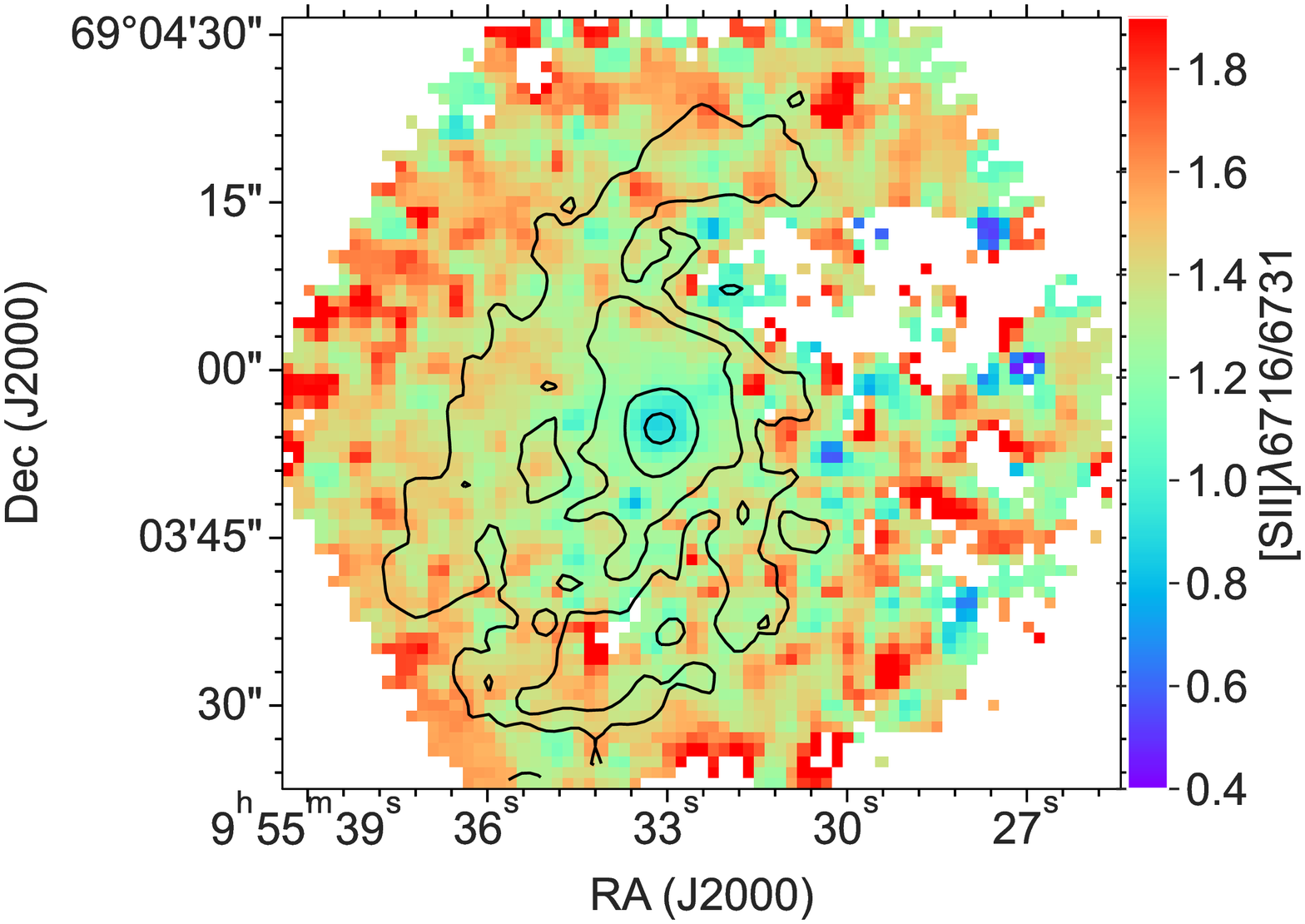}
\includegraphics[width= 3.7in]{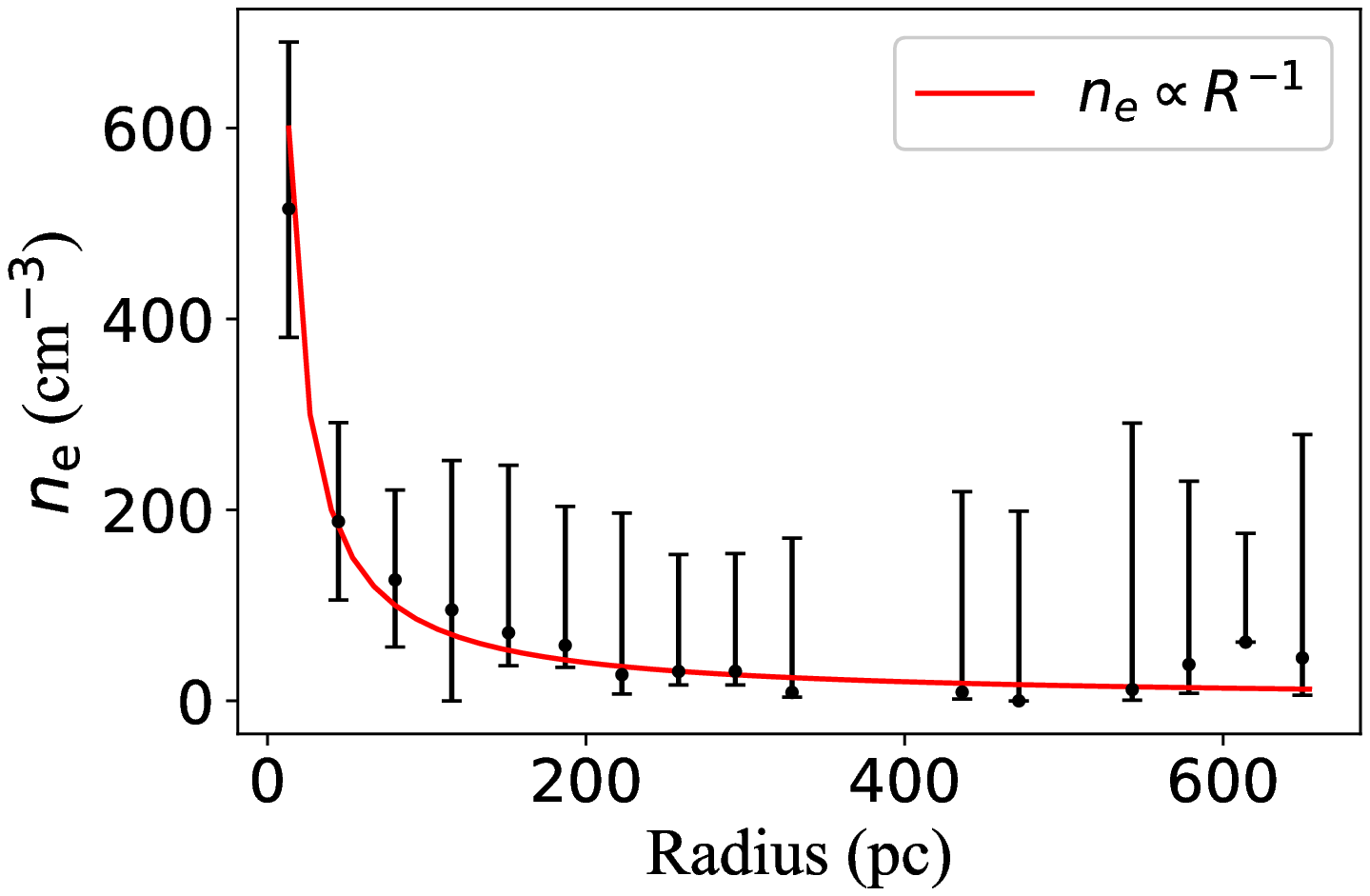}
\caption{$Left$: Line ratio of the [S\,{\sc ii}] doublet overlaid with HST H$\alpha$ intensity contours. $Right$: The azimuthally-averaged radial distribution of electron density inferred from the [S\,{\sc ii}] ratio, which is compared with a $R^{-1}$ distribution. Since the electron density conversion is only valid for [S\,{\sc ii}] ratios between 0.4 -- 1.4, pixels and bins with ratios outside this range have been excluded.
\label{fig:sii}}
\end{figure}

The line ratio between the [S\,{\sc ii}] doublet, [S\,{\sc ii}]$\lambda$6716/[S\,{\sc ii}]$\lambda$6731, is sensitive to electron density within the range of 10$^{1-4}$ cm$^{-3}$. The [S\,{\sc ii}] line ratio map is presented in the left panel of Figure \ref{fig:sii}, from which the electron density distribution can be determined, adopting the relation from \cite{2014A&A...561A..10P} and assuming an electron temperature of $T\rm_e = 10000~K$. 
The line ratio is the lowest in the nucleus, corresponding to an electron density of $\sim$ 600 cm$^{-3}$, which is consistent with the estimation in \cite{2011MNRAS.413..149S}. 
Relatively low ratios (high electron densities) are also found in regions coincident with the nuclear spiral. 
Otherwise the line ratio generally increases outward, corresponding to a decreasing trend of electron density until $R \sim$ 300 pc, beyond which the line ratio appears flat and is no longer sensitive to the electron density.
The azimuthally-averaged radial distribution of the electron density is shown in the right panel of Figure \ref{fig:sii}, which can be roughly described by a $R^{-1}$ profile. 
\\


\section{Discussion} \label{sec:4}



In the previous sections, we have presented the basic properties of one of the nearest LINERs in a spatially-resolved fashion. 
The ionized gas mass of this LINER can be estimated based on the observed H$\alpha$ luminosity \citep{Kurt_2012}, $M\rm_{ion} = \it \mu m\rm_p \it L\rm_{H\alpha}/\it \gamma$, where $\mu = 1.4$ is the mass per H atom, $m_p$ is the proton mass, $L\rm_{H\alpha}$ is the H$\alpha$ luminosity, $n_e$ is the electron density and $\gamma$ is the effective volume emissivity, which is $3.56\times10^{-25}$ erg cm$^3$ s$^{-1}$ in Case B recombination \citep{1989agna.book.....O}. 
The ionized gas mass within the central 500 pc in radius is thus estimated to be $\sim 2.2 \times 10^5 ~\rm M_\odot$, taking the radial distribution of electron density from Figure~\ref{fig:sii}. 
We note that the majority of this mass is found in the core and the nuclear spiral, but a non-negligible fraction of the ionized gas exists in a more diffuse manner. 
This mass is to be contrasted with the total molecular gas mass of $\sim 4.3\times 10^6 ~\rm M_\odot$ within the central kpc based on CO lines \citep{2007A&A...473..771C}. 
From the equivalent hydrogen column density inferred from the dust extinction map, which is concentrated in the nuclear spiral (Figure~\ref{fig:av}), we can also estimate a mass $\sim 1.0\times 10^6 ~\rm M_\odot$ in the dusty cold gas. 
Therefore the ionized gas accounts for a few percent of the circumnuclear cold/warm ISM. Considering also the fact that the emission lines and the dust extinction are highly co-spatial, it is reasonable to suggest a geometry in which the emission lines primarily arise from the ionized outer layer of the dusty cold gas.
\\

It is interesting to compare with the case of M31, in which a nuclear spiral of similar physical extent has also been identified in both ionized gas and dusty cold gas \citep{2009MNRAS.397..148L}. However, the mass of the ionized gas in the central kpc of M31 is found to be a few $10^3~\rm M_\odot$ \citep{1985ApJ...290..136J, 2009MNRAS.397..148L}, nearly two orders of magnitude lower than estimated for M81 here. In the mean time, the circumnuclear cold gas in M31, primarily in the form of molecular gas, has a mass of a few $10^6~\rm M_\odot$ \citep{2019MNRAS.484..964L}, similar to the case of M81. The large difference between the ionized gas-to-molecular gas mass ratio might be attributed to the presence (absence) of an irradiating LLAGN in M81 (M31). 
\\

From the velocity field of both H$\alpha$ and [O\,{\sc iii}], we have tentatively identified a bi-conical outflow of ionized gas in the core region (Figure~\ref{fig:v}). 
The outflow rate can be estimated following \cite{Kurt_2012}:
\begin{equation}
\dot{M}_{\rm out}=\frac{\mu m\rm_p \it L\rm_{H\alpha} \it v\rm_{\rm out}}{\gamma \it n\rm_e \it r\rm_{\rm out}}
\end{equation}
where $v_{\rm out}$ is the outflow velocity and $r_{\rm out}$ is the outflow radius. For the bi-conical outflow outlined in Figure \ref{fig:v}, we adopt $n\rm_e = 200$ cm$^{-3}$, $v\rm_{out} = 50/cos(\it i)$ km s$^{-1}$ (since the outflow is likely along the disk plane, the velocity is corrected for an inclination angle $i = 58^\circ$) and an outflow radius of 10$\arcsec$ ($\sim$180 pc), which results in $\dot{M}_{\rm out} \approx 1.4 \times 10^{-3}$ M$_\odot$ yr$^{-1}$. 
Perhaps coincidentally, this outflow rate is comparable to the estimated inflow rate ($\sim4\times 10^{-3}$ M$_\odot$ yr$^{-1}$) of ionized gas detected in the central few arcsec ($\sim$50 pc) based on the GMOS observations \citep{2011MNRAS.413..149S}. 
Interestingly, it is also comparable to the outflow rate ($\sim2\times 10^{-3}$ M$_\odot$ yr$^{-1}$) of a parsec-scale hot wind recently detected by \citet{2021NatAs...5..928S} based on {\it Chandra} X-ray observations. This hot wind is presumably driven by the hot accretion flow onto M81*, manifesting itself as a hot ($\sim 10^8$ K) plasma with a line-of-sight velocity of $\sim$3000 km s$^{-1}$.
It is plausible that the ionized gas outflow at the hundred parsec-scale is associated with the parsec-scale hot wind. 
For instance, when the hot wind propagates outward, due to its large opening angle \citep{2021NatAs...5..928S}, pre-existed gas in the nuclear spiral can be entrained into a decelerating outflow. Such a case would have important implications on the feedback mechanism of LLAGNs, which have gained growing attention in recent years \citep{2016Natur.533..504C, 2019ApJ...885...16Y, 2017MNRAS.465.3291W,  2021NatAs...5..928S}. 
The physical reality of the bi-conical outflow and its relation to the hot wind are certainly worthy of further exploration, e.g., with higher-resolution spectroscopic observations in and around the core.  
\\


The spatially-resolved BPT and WHAN diagrams (Figure~\ref{fig:ratio}) demonstrate that the circumnuclear region of M81, in particular the central $\sim$ 150 pc, is a LINER. 
However, a substantial radial gradient exists in the line intensities (Figure~\ref{fig:flux}), and  significant spatial variations are clearly present in the [O\,{\sc iii}]/H$\beta$ ratio (Figure~\ref{fig:r2}).
This implies that the ionization/excitation mechanism of this region is more complicated than the naive case of photoionization by a central source, although the LLAGN, M81*, must be playing a significant, if not dominant, role.   
Hot evolved stars such as pAGB stars are arguably the most viable candidate, after a LLAGN, to account for extended low-excitation lines \citep{1994A&A...292...13B, 2013A&A...558A..43S}. 
The extensive multi-wavelength observations available for the inner region of M81 makes it particularly suited for  a quantitative examination of the pAGB scenario, as well as other secondary, but perhaps ubiquitous, ionization/excitation mechanisms, such as shocks and cosmic-rays.  
Also intriguing is the possibility that M81* had been significantly more active and luminous in the recent past, as suggested by, e.g.,  \citet{2007A&A...463..551S}. 
The light crossing time of the PPAK FoV is only $\sim10^3$ yr, which is much shorter than the recombination time of $\sim10^4$ yr for a typical electron density of 10 cm$^{-3}$. 
Delayed recombination and de-excitation of the circumnuclear gas after the passing of the light front from the AGN phase could lead to the Seyfert signature tentatively seen at large radii, a possibility that deserves further investigation. 
A more quantitative analysis of the competing ionization mechanisms constrained by the spatially-resolved spectroscopy of PPAK will be presented in a subsequent work (Z. Li et al. in preparation). 
\\

\section{Summary} 
\label{sec:sum}
We have carried out the first optical IFS observation dedicated for the central kpc of M81, utilizing the CAHA 3.5 m telescope. In this work, we focus on the ionized gas properties from the analysis of the emission lines. 
Our main results can be summarized as follows:

\begin{itemize}
	\item The optical emission lines are strongest in the inner $\sim$ 100 pc core region as well as in the nuclear spiral. In addition, a more diffuse component of ionized gas is detected throughout the central kpc. 
	
	\item 
	Dust extinction features revealed by a high-resolution, broad-band HST image exhibits a highly similar morphology with the nuclear spiral, suggesting that much of the ionized gas and dusty cold gas are co-spatial. 
	
	\item The velocity field of ionized gas reveals a pair of arc-shaped features roughly symmetric about the nucleus at distances of $\sim 120-250$ pc, which might be tracing a bi-conical outflow along the disk.
	The mass outflow rate is inferred to be order $10^{-3}$ M$_\odot$ yr$^{-1}$, implying for a significant means of mechanical power injected into the circumnuclear region. 
	
	\item A large fraction of the circumnuclear region of M81 can be classified as a LINER, base on the BPT and WHAN diagrams. However, substantial spatial variations in the line intensities and line ratios suggest that different ionization/excitation mechanisms, rather than just a central source of photoionization, may conspire to produce the observed signatures.  
\end{itemize}

Future work of detailed photoionization modeling and stellar population analysis promise to obtain a comprehensive picture of the circumnuclear region in M81, 
and will help advance our general understanding of the interplay between SMBHs and their immediate environments.
\\


\begin{acknowledgements}

This research has made use of data and software provided by the Calar Alto Observatory.	
Z.N.L. and Z.Y.L. acknowledge support by the National Key Research and Development Program of China (grant 2017YFA0402703), the National Natural Science Foundation of China (grant 11873028), and the science research grants from the China Manned Spaced Project (CMS-CSST-2021-B02). R.G.B. acknowledges financial support from the State Agency for Research of the Spanish MCIU through the “Center of Excellence Severo Ochoa” award to the Instituto de Astrof\'isica de Andaluc\'ia (SEV-2017-0709) and  P18-FRJ-2595 (Junta de Andaluc\'{\i}a).
S.F. acknowledges support by National Natural Science Foundation of China (No. 12103017), Natural Science Foundation of Hebei Province (No. A2021205001) and Science Foundation of Hebei Normal University (No. L2021B08). 
We thank Yanmei Chen and Lei Hao for helpful discussions.

\end{acknowledgements}

\bibliographystyle{aasjournal.bst}

\bibliography{m81}

\appendix

The intensity maps of [O\,{\sc ii}]$\lambda$3727 and [S\,{\sc ii}]$\lambda$6716 are presented in Figure \ref{fig:A1}, which share a similar overall morphology with the [O\,{\sc ii}] and H$\alpha$ intensity maps.

\setcounter{figure}{0}
\renewcommand\thefigure{A\arabic{figure}}
\begin{figure}
\centering
\includegraphics[width=3.5in]{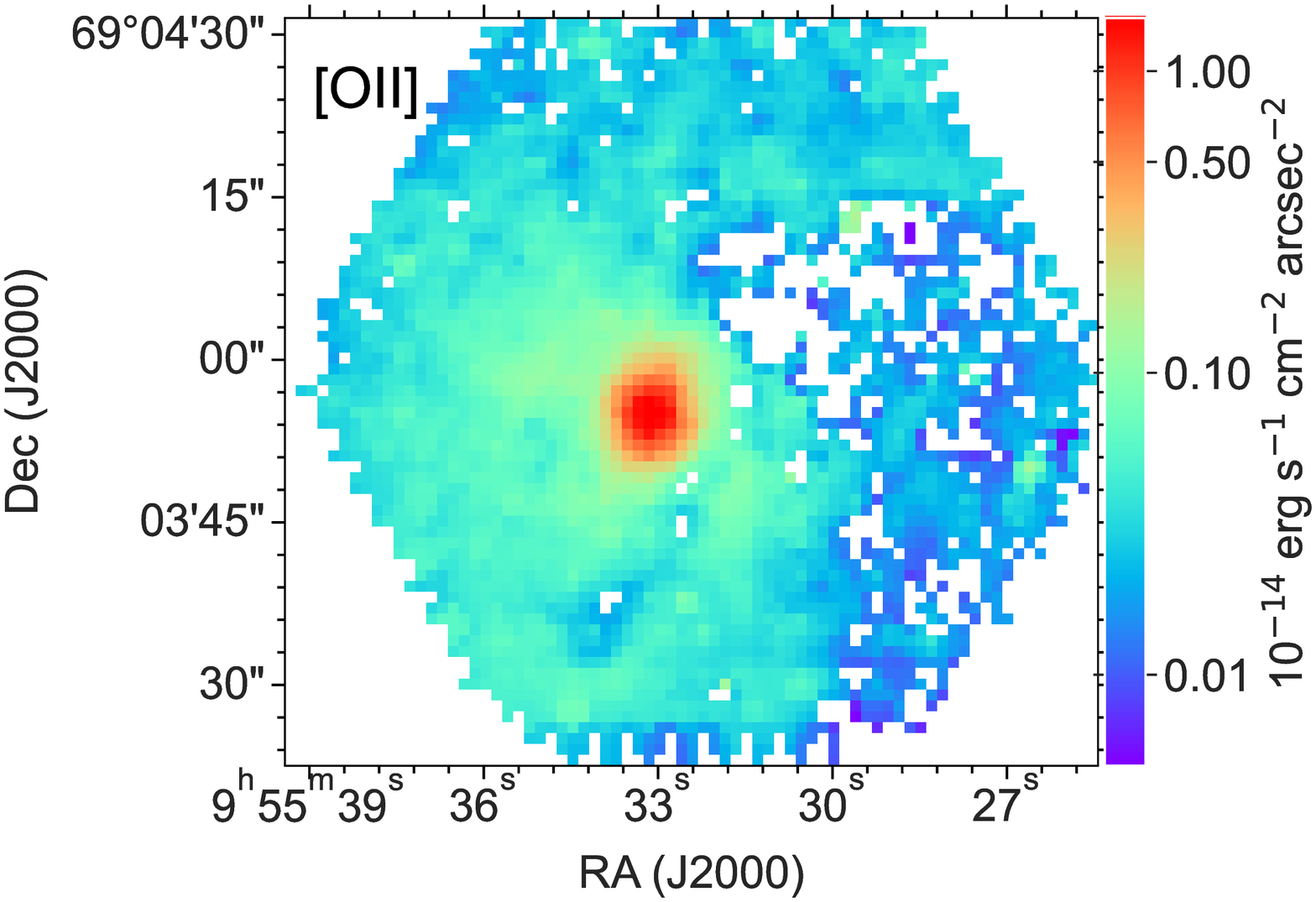}
\includegraphics[width=3.5in]{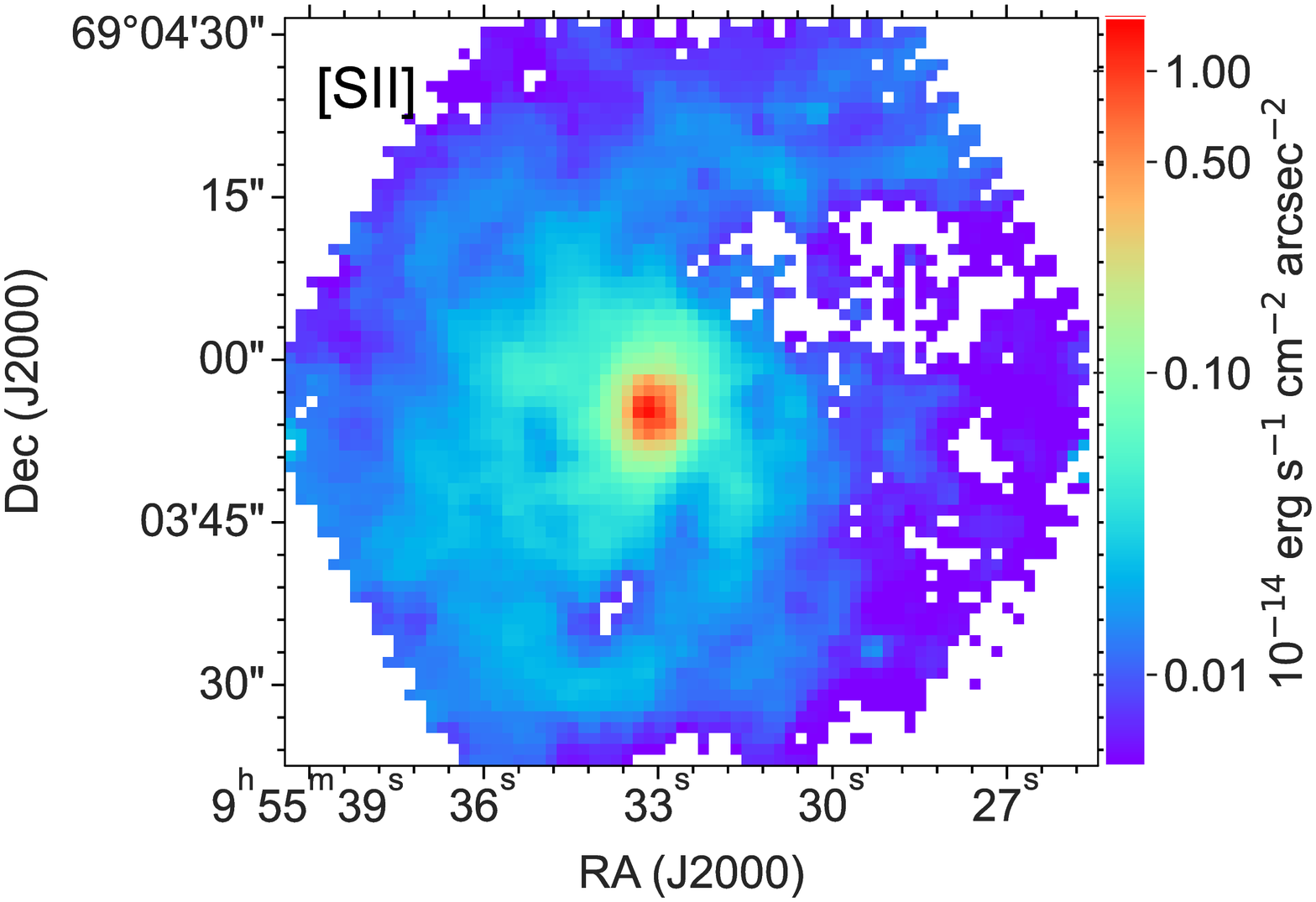}
\caption{Surface brightness distribution of the [O\,{\sc ii}]$\lambda$3727 ($left$) and [S\,{\sc ii}]$\lambda$6716 ($right$) emission lines. Pixels with $S/N < 5$ have been masked. \label{fig:A1}}
\end{figure}

\end{document}